\documentclass[11pt,onecolumn,draftclsnofoot,letterpaper]{IEEEtran}
\usepackage{graphicx}
\usepackage{subfigure}
\usepackage{epsfig}
\usepackage{amssymb}
\usepackage{amsmath}
\usepackage{amsfonts}

\newcommand{\va}{\mbox{${\bf a}$}}

\newcommand{\vp}{\mbox{${\bf p}$}}

\newcommand{\vx}{\mbox{${\bf x}$}}

\newcommand{\vt}{\mbox{${\bf t}$}}
\newcommand{\vw}{\mbox{${\bf w}$}}

\newcommand{\vs}{\mbox{${\bf s}$}}

\newcommand{\vu}{\mbox{${\bf u}$}}
\newcommand{\vy}{\mbox{${\bf y}$}}

\newcommand{\vn}{\mbox{${\bf n}$}}

\newcommand{\mB}{\hbox{{\bf B}}}

\newcommand{\mC}{\hbox{{\bf C}}}

\newcommand{\mH}{\hbox{{\bf H}}}

\newcommand{\mR}{\mbox{{$\bf R$}}}

\newcommand{\ga}{\alpha}
\newcommand{\gb}{\beta}

\newcommand{\gr}{\rho}
\newcommand{\gs}{\sigma}

\newcommand{\gD}{\Delta}

\def\bm#1{\mbox{\boldmath $#1$}}
\newcommand{\vga}{\mbox{$\bm \alpha$}}

\newcommand{\cR}{{\cal R}}

\newtheorem{theorem}{Theorem}[section]
\newtheorem{lemma}[theorem]{Lemma}

\newtheorem{prop}{Proposition}[section]
\newtheorem{claim}{Claim}[section]

\newtheorem{definition}{Definition}[section]
\newtheorem{question}{Question}[section]
\newtheorem{coro}{Corollary}[section]

\newcommand{\beq}{\begin{equation}}
\newcommand{\eeq}{\end{equation}}
\newcommand{\bea}{\begin{array}}
\newcommand{\ena}{\end{array}}
\newcommand{\bds}{\begin {itemize}}
\newcommand{\eds}{\end {itemize}}
\newcommand{\bdf}{\begin{definition}}
\newcommand{\blm}{\begin{lemma}}
\newcommand{\edf}{\end{definition}}
\newcommand{\elm}{\end{lemma}}
\newcommand{\bthm}{\begin{theorem}}
\newcommand{\ethm}{\end{theorem}}
\newcommand{\bprp}{\begin{prop}}
\newcommand{\eprp}{\end{prop}}
\newcommand{\bcl}{\begin{claim}}
\newcommand{\ecl}{\end{claim}}
\newcommand{\bcr}{\begin{coro}}
\newcommand{\ecr}{\end{coro}}
\newcommand{\bquest}{\begin{question}}
\newcommand{\equest}{\end{question}}

\newcommand{\rarrow}{{\rightarrow}}

\renewcommand{\baselinestretch}{1.25}
\begin{document}
\title{Game theory and the frequency selective interference channel - A tutorial}
\author{Amir Leshem$^{1,2}$  and Ephraim Zehavi$^1$ \thanks{
$^1$School of Engineering, Bar-Ilan University, Ramat-Gan,
52900, Israel. $^2$Faculty of EEMCS, Delft University of Technology. This work was supported by Intel Corporation and by the Netherlands
Foundation of Science and Technology. e-mail: leshema@eng.biu.ac.il .}
}
\date{\today}
\maketitle
\begin{abstract}
This paper provides a tutorial overview of game theoretic techniques used for communication over
frequency selective interference channels. We discuss both
competitive and cooperative techniques.

\end{abstract}
\section{Communication over interference limited channels}
The success of unlicensed broadband communication has led to very rapid deployment of
communication networks that work independently of each other using a relatively narrow
spectrum. For example the 802.11g standard is using the ISM band which has a total bandwidth of 60
MHz. This band is divided into 12 partially overlapping bands of 20 MHz. The success of these technologies might
become their own limiting factor. The relatively small number of channels and the massive use of the
technology in densely populated urban metropolitan areas will cause significant mutual interference.
This is especially important for high quality real time distribution of multi-media content that is
intolerant to errors as well as latency. Existing 802.11 (WiFi) networks have very limited means to
coordinate spectrum with other interfering systems. It would be highly desirable to improve the interference
environment by distributed spectral coordination between the different access points.
Another scenario is that of centralized access points such as 802.16 (WiMax) where the resources are allocated
centrally by a single base station. Similar situation is facing the advanced DSL systems such as ADSL2+ and VDSL.
These systems are currently limited by crosstalk between the lines. As such the DSL environment is another example
of highly frequency selective interference channel.
While the need to operate over interference limited frequency selective channels is clear in many of
the current and future communication technologies, the theoretical situation is much less satisfying.
The capacity region of the interference channel is still open (see \cite{meulen94} for short overview) even for the
fixed channel two user case. In recent years great advances in understanding the situation for flat channels under weak
interference have been achieved. It can be shown that in this case treating the interference as noise is almost
optimal. On the other hand for the medium strength interference as is typical in the wireless environment, the
simplest strategy is by using orthogonal signaling, e.g. TDMA/FDMA, for high spectral efficiency networks, or CDMA for very strong interference
with low spectral efficiency per user.
Moreover, sequential cancelation techniques that are required for the best known capacity region in the medium interference case \cite{han81}
are only practical for small number of interferers. The interference channel is a conflict situation, and not every achievable rate pair
(from informational point of view) is actually reasonable operating point for the users. This conflict is the reason
to analyze the interference channel using game theoretical tools. Much work has been done on competitive game theory
applied to frequency selective interference channel, with the early works of Yu et al. \cite{yu2002} and subsequent works of
Scutari et al. (see \cite{scutari2008} and the references therein). A particularly interesting topic is the use of  generalized Nash
games to the weak interference channel \cite{pang_information_theory}, and the algorithm in \cite{noam2009} which extends the FM-IWF to iterative pricing
under fixed rate constraint.

The fact that competitive strategies can result in significant degradation due to the prisoner's dilemma has been called the price of anarchy
\cite{papadimitriou2001}.  In the interference channel case, a simple case has been analyzed by Laufer and Leshem \cite{laufer2005},
who characterized the cases of prisoner's dilemma in interference limited channels. To overcome the sub-optimality of the competitive approach
we have two alternatives: Using repeated games or using cooperative game theory. Since most works on repeated games concentrated on flat fading
channels, we will mainly concentrate on cooperative game theoretic approaches. One of the earliest solutions for cooperative games is the
Nash bargaining solution \cite{nash50}. Many papers in recent years were devoted to analyzing the Nash bargaining solution for
the frequency flat interference channel in the SISO \cite{leshem2006}, \cite{boche07}, MISO \cite{jorswieck2008}, \cite{gao2008} and
MIMO cases \cite{nokleby07}. Interesting extensions for log-convex utility functions appeared in \cite{schubert2008}. Another interesting application
of the bargaining techniques discussed here is for multimedia distribution networks. Park and Van der Schaar
\cite{park2007} used both Nash bargaining and the generalized Kalai-Smorodinski solutions \cite{kalai556} for multimedia resource management.
Another alternative cooperative model was explored in \cite{mathur2006} where, the cooperation between rational wireless players
was studied using coalitional  game theory by allowing the receivers to cooperate using joint decoding.

In the context of frequency selective interference channels much less research has been done. Han et al. \cite{han2005}
in a pioneering work, studied the Nash Bargaining under FDM/TDM strategies and total power constraint. Unfortunately, the algorithms proposed
were only sub-optimal. Iterative sub-optimal algorithms to achieve Nash
bargaining solution for spectrum  allocation under average power constraint have been applied in \cite{zhaoyi2008}.
Only recently, we have managed to overcome the difficulties by imposing a total PSD mask constraint \cite{leshem2008},
in order to obtain computationally efficient solutions to the bargaining problem in the frequency selective SISO and MIMO cases under TDM/FDM strategies.
Furthermore, it can be shown \cite{zehavi2009} that the PSD limited case can be used to derive a computationally efficient converging algorithm
also in the total power constraint case. A very interesting problem is allowing the users to treat the interference as noise in some bands, while
using orthogonal FDM/TDM strategies in others. This is a very
challenging task, since the Nash bargaining solution involves a highly non-convex power allocation problem, instead of the simpler orthogonal signaling.

As discussed before, the frequency selective interference channel is very interesting both from practical point of view and from information theoretic
point of view. We can show that it has many interesting aspects from game theoretic point of view, and that various levels of interference admit
different types of game theoretic technique. The purpose of this paper is to provide an overview of the various game theoretic techniques involved
in the analysis and algorithms used for frequency selective interference channels. We
demonstrate how game theory can be applied in specific scenarios and discuss signal processing aspects of the required game theoretic solutions.
Our main goal is to discuss in a tutorial manner some new applications of game theory to frequency selective interference channels. First, we
outline various interference channel scenarios with emphasis on
frequency selective channels. Then we discuss the basic concepts of game
theory required for analyzing these channels: Nash equilibrium, strategies, Generalized Nash games and Nash bargaining theory.
Using these concepts we discuss the specific translations into working strategies for the communication models and discuss important signal
processing aspects such as spectrum sensing and centralized and distributed strategies.
To wrap up the discussion with real life applications, end up with two case studies:
DSL and WiMax where we demonstrate the gains of the techniques on real channels. We end up with conclusions and future research directions.

\section{Introduction to interference channels}
\label{interference_channel_model}
Computing the capacity region of the interference channel is an open problem in information
theory \cite{cover}. A good overview of the results until 1985 is given by van der Meulen  \cite{meulen94}
and the  references therein. The capacity region of general interference channel
is not known yet. However, in the last forty five years of research some
progress has been made.
The best known achievable region for the general
interference channel is due to Han and Kobayashi  \cite{han81}. The
computation of the Han and Kobayashi formula for a general discrete memoryless
channel is in general too complex. Recently large advance in obtaining upper bounds on the rate region have been obtained
especially for the case of weak interference.

A 2x2 Gaussian interference channel
in standard form (after suitable normalization) is given by:
\beq
\label{standard_IC}
\vx=\mH \vs +\vn, \quad
\mH=\left[
\bea{cc}
1 & \ga_1 \\
\ga_2 & 1
\ena
\right]
\eeq
where, $\vs=[s_1,s_2]^T$, and $\vx=[x_1,x_2]^T$ are sampled values of the input and
output signals, respectively. The noise vector $\vn$ represents the
 additive Gaussian noises with zero mean and unit variance. The powers of the
input signals are constrained to be less than $P_1,P_2$, respectively. The off-diagonal elements of $\mH$, $\ga_1,\ga_2$ represent the degree of
interference present. The major difference between the interference channel and the multiple access channel is that both encoding and
decoding of each channel are performed separately and independently, with no information sharing between receivers.

The capacity region of the Gaussian interference channel with very
strong interference (i.e., $\ga_1 \ge 1+P_1$, $\ga_2 \ge 1+P_2$ )
is given by \cite{carleil78}
\beq
\label{VSI_RR}
R_i \le \log_2(1+P_i), \ \ i=1,2.
\eeq
This surprising result shows that very strong interference dose not reduce the capacity of the users.
A Gaussian interference channel is said to have strong interference if  $\min\{\ga_1,\ga_2\}>1$. Sato
\cite{sato81} derived an achievable capacity region (inner bound) of Gaussian
interference channel as intersection of two multiple access Gaussian capacity
regions embedded in the interference channel. The achievable region is the
intersection of the rate pair  of the rectangular region of the very strong
interference (\ref{VSI_RR}) and the following region:
\beq
R_1+R_2 \le \log_2\left(\min\left\{1+P_1+\ga P_2,1+P_2+\gb P_1\right\} \right).
\eeq

While the two user flat interference channel is a well studied (although not solved) problem, much less is known in the
frequency selective case. An  $N\times N$ frequency selective Gaussian
interference channel is given by:
\beq
\label{standard_IC_matrix}
\bea{c}
\vx_k=\mH_k \vs_k +\vn_k
\qquad k=1,...,K \\
\mH_k=\left[
\begin{array}{ccc}
             h_{11}(k) & \hdots &  h_{1N}(k)\\
             \vdots & \ddots & \vdots \\
             h_{N1}(k) & \hdots &  h_{NN}(k)\\
\end{array}
\right].
\ena
 \end{equation}
where, $\vs_k$, and $\vx_k$ are sampled values of the input and output signal vectors at
frequency $k$, respectively. The noise vector $\vn_k$ represents an
 additive white Gaussian noise with zero mean and unit variance. The power spectral density (PSD)
of the input signals are constrained to be less than $p_1(k),p_2(k)$
respectively. Alternatively, only a total power constraint is given. The off-diagonal elements of $\mH_k$, represent the
degree of interference present at frequency $k$. The main difference between interference channel
and a multiple access channel (MAC) is that in the interference channel, each component
of $s_k$ is coded independently, and each receiver has access to a single element of
$\vx_k$. Therefore, iterative decoding schemes are much more limited, and typically
impractical for large number of users.

\begin{figure}
    \begin{center}
    \subfigure[Interference channel]
    {
    \label{interference_channel}
    \includegraphics[width=0.3\textwidth]{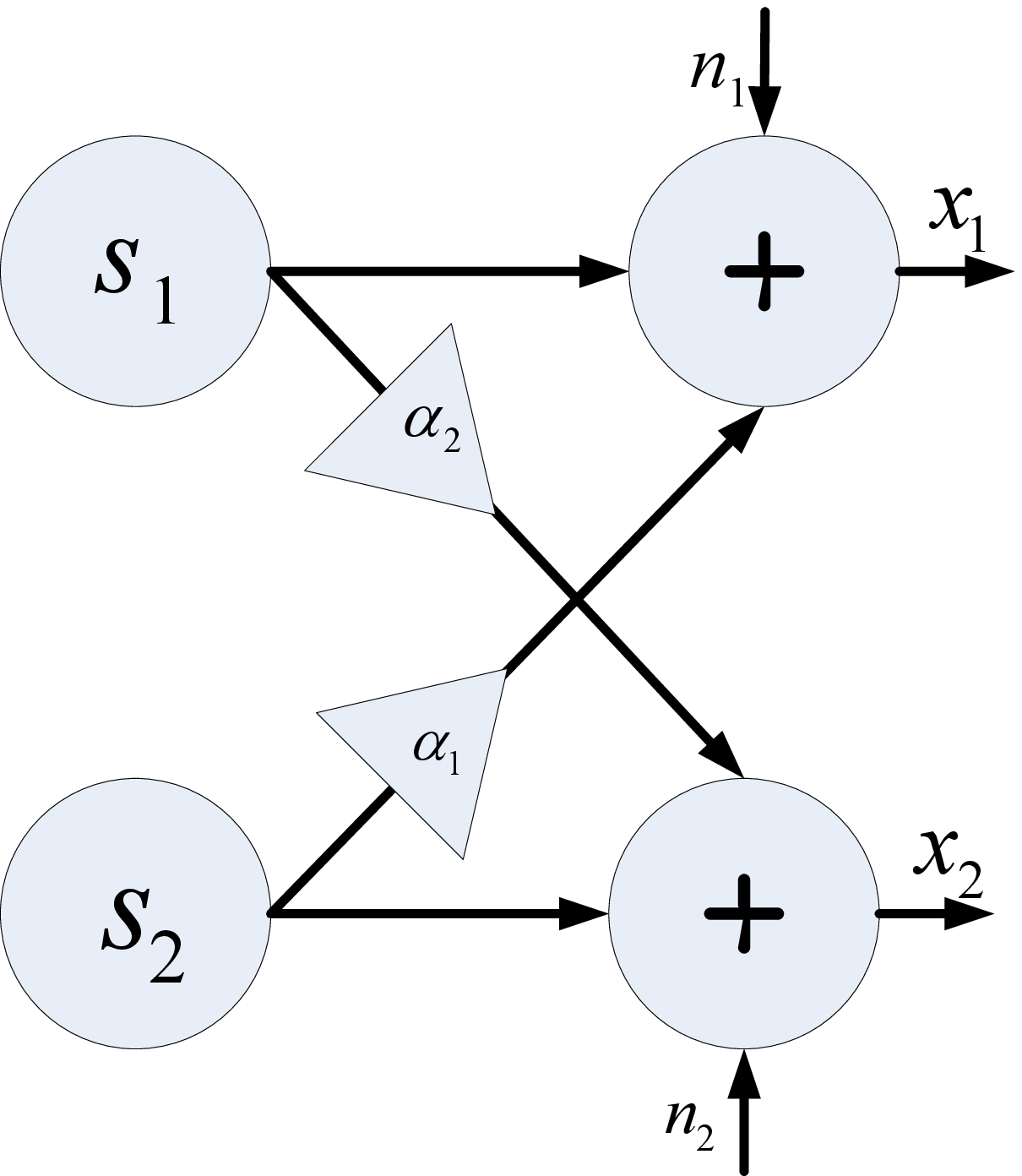}}
    \subfigure[TDMA and joint TDMA/OFDMA]
    {
    \label{ofdma}
    \includegraphics[width=0.35\textwidth]{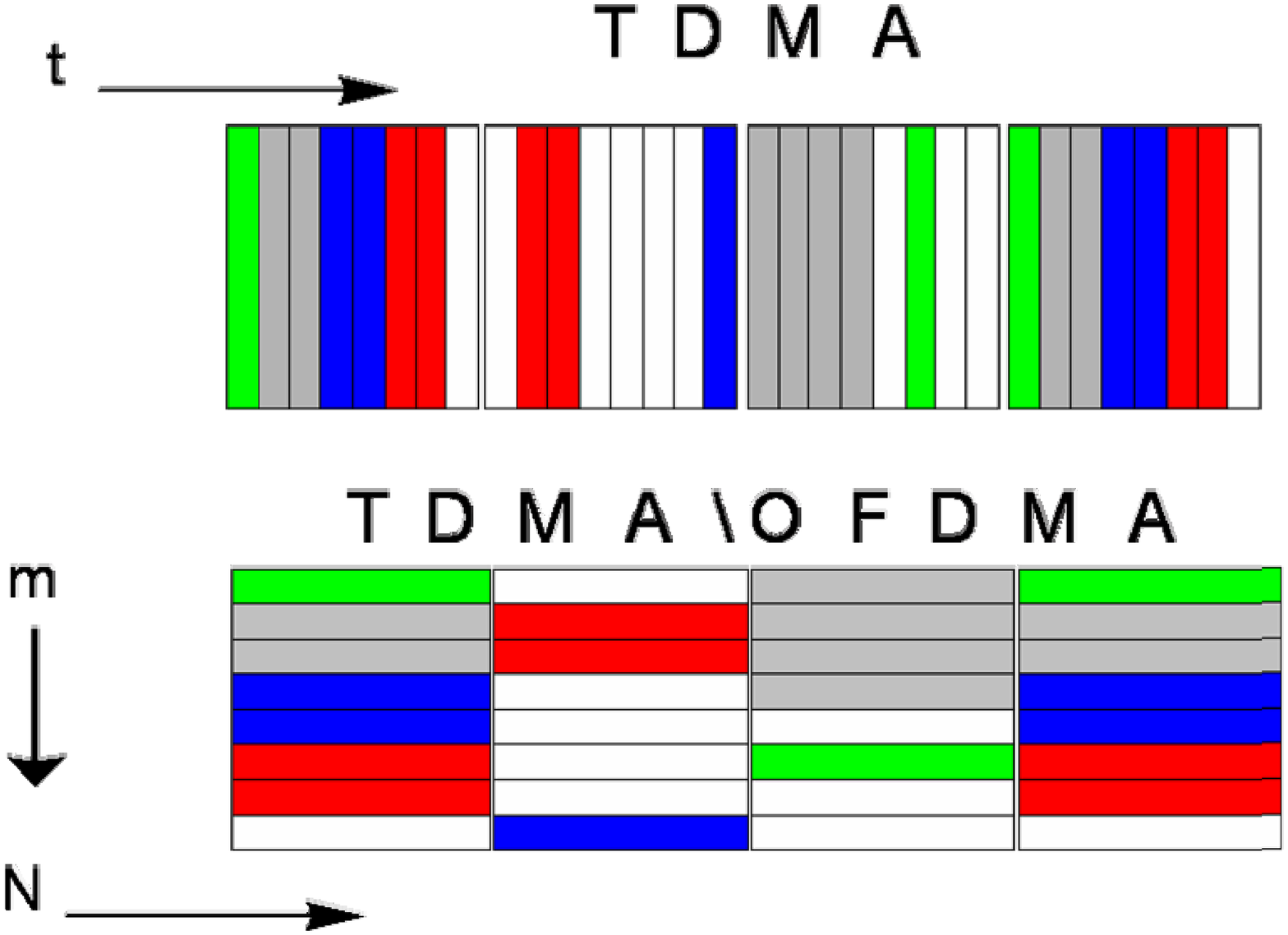}}
    \end{center}
    \caption{(a) Standard form interference channel. (b) TDMA and joint TDMA/OFDMA}
\end{figure}

To overcome this problem there are two simple strategies. When the interference is sufficiently weak, the common wisdom
is to treat the interference as noise, and code at a rate corresponding to the total noise. When the interference is stronger, i.e,
Signal to Interference Ratio (SIR) is significantly lower than Signal to additive Noise Ratio (SNR), treating the interference as noise can be
highly inefficient. One of the simplest ways to deal with medium to strong interference channels is
through orthogonal signaling. Two extremely simple orthogonal schemes are using
FDM or TDM strategies. These techniques allow a single user detection (which will be assumed throughout
this paper) without the need to complicated multi-user detection.
The loss of these techniques compared to techniques requiring joint decoding has
been thoroughly studied, e.g., \cite{carleil78} in the constant channel case, showing degradation compared to the techniques requiring
joint or sequential decoding. However, the widespread use of FDMA/TDMA as well as collision avoidance medium
access control (CSMA) techniques, make the analysis of these techniques very important from practical point
of view as well. For frequency selective channels (also known
as ISI channels) we can combine both strategies by allowing time
varying allocation of the frequency bands to the different users as shown in figure \ref{ofdma}.

In this paper we limit ourselves to joint FDM and TDM scheme where an
assignment of disjoint portions of the frequency band to the several
transmitters is made at each time instance. This technique is widely
used in practice because simple filtering can be used at the
receivers to eliminate interference.
All of these schemes as to operate under physical and regulation constraint like Average power constraint  or/and PSD  mask constraint.
%
\section{Basic concepts of cooperative and competitive game theory}
\label{game_theory_concepts}
In this section we review the basic concepts of game theory in an abstract setting. Our focus is on concepts that
have been found to be relevant to the frequency selective interference channel. We begin with competitive game theory and
then continue to describe the cooperative solutions. The reader is referred to the excellent books of \cite{owen}, \cite{basar82}
and \cite{osborn} for more details and for proofs of the main results mentioned here.

\subsection{Static competitive games and the Nash equilibrium}
An static $N$ player game in strategic form is a three tuple $\left(N,A,\vu\right)$ composed of a set of players $\left\{1,...,N\right\}$, a
set of possible combinations of actions of each player denoted by $A=\prod_{n=1}^N A_n$, where $A_n$ is the set of actions for the $n$'th player
and a vector utility function $\vu=\left[u_1,...,u_N \right]$, where
$u_n(a_1,...,a_N):\prod_{n=1}^N A_n \rarrow \cR$ is the utility of the $n$'th player when strategy vector $\va=(a_1,...,a_N)$ has been played.
The interpretation of $u_n$ is that player $n$ receives a payoff of
 $u_n(a_1,...,a_N)$ when the players have chosen actions $a_1,...,a_N$. The game is finite when for all $n$, $A_n$ is a finite set.

 A special type of competitive games are the constant sum games. A game is constant sum, if for all action vectors $\va$,
 $\sum_{n=1}^N u_n(\va)=c$ for some constant $c$. When the game is constant sum we can subtract $c/N$ from each utility and obtain a zero sum game
 that has the same properties as the original game. A two players zero sum game is strictly competitive since anything gained by one player leads
 to a loss to the other player.

\begin{figure}
    \begin{center}
    \subfigure[Multiple Access game]
    {
    \label{mac_game}
    \includegraphics[width=0.4\textwidth]{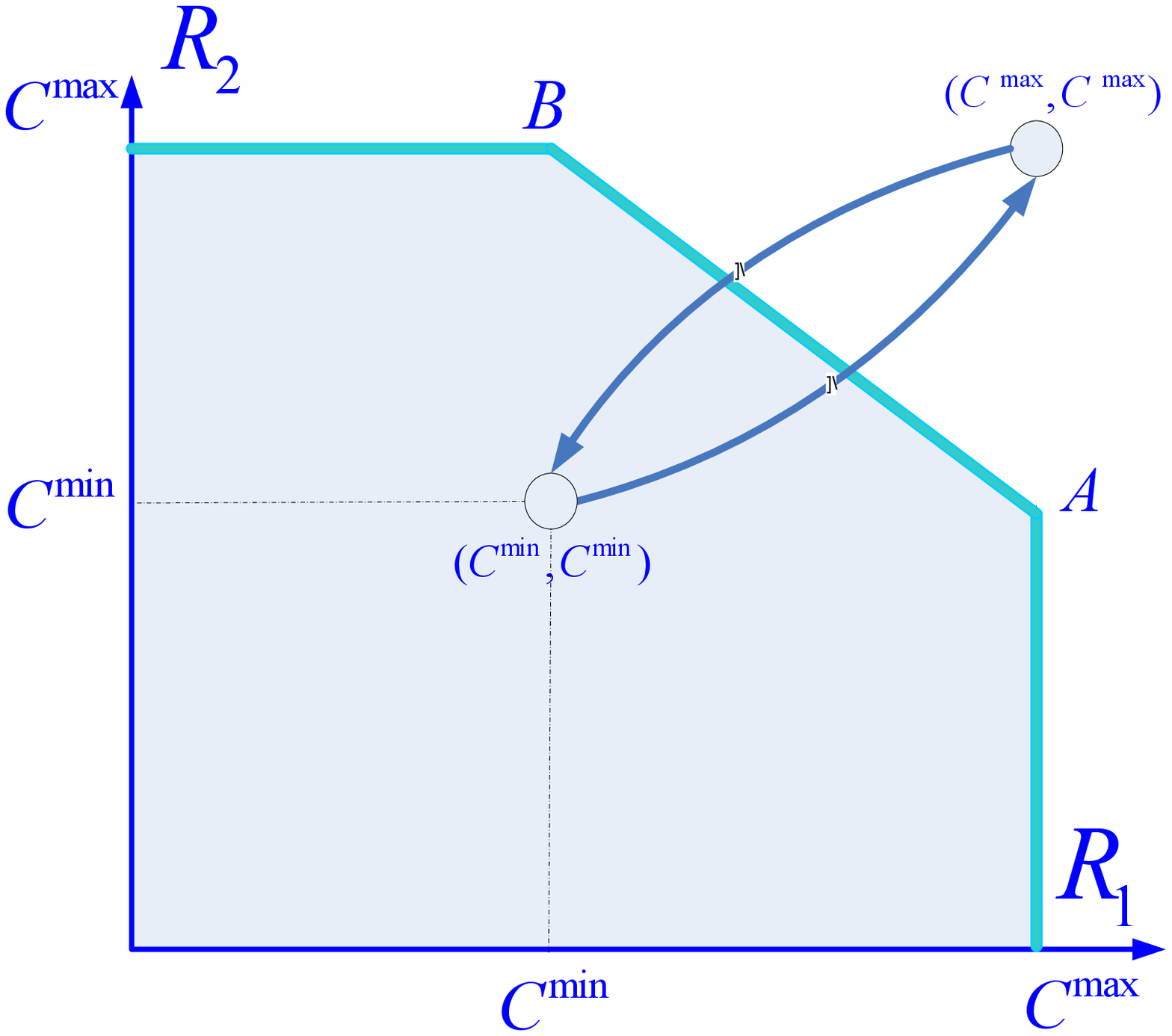}}
    \subfigure[Prisoner's dilemma regions]
    {
    \label{pd_region}
    \includegraphics[width=0.4\textwidth]{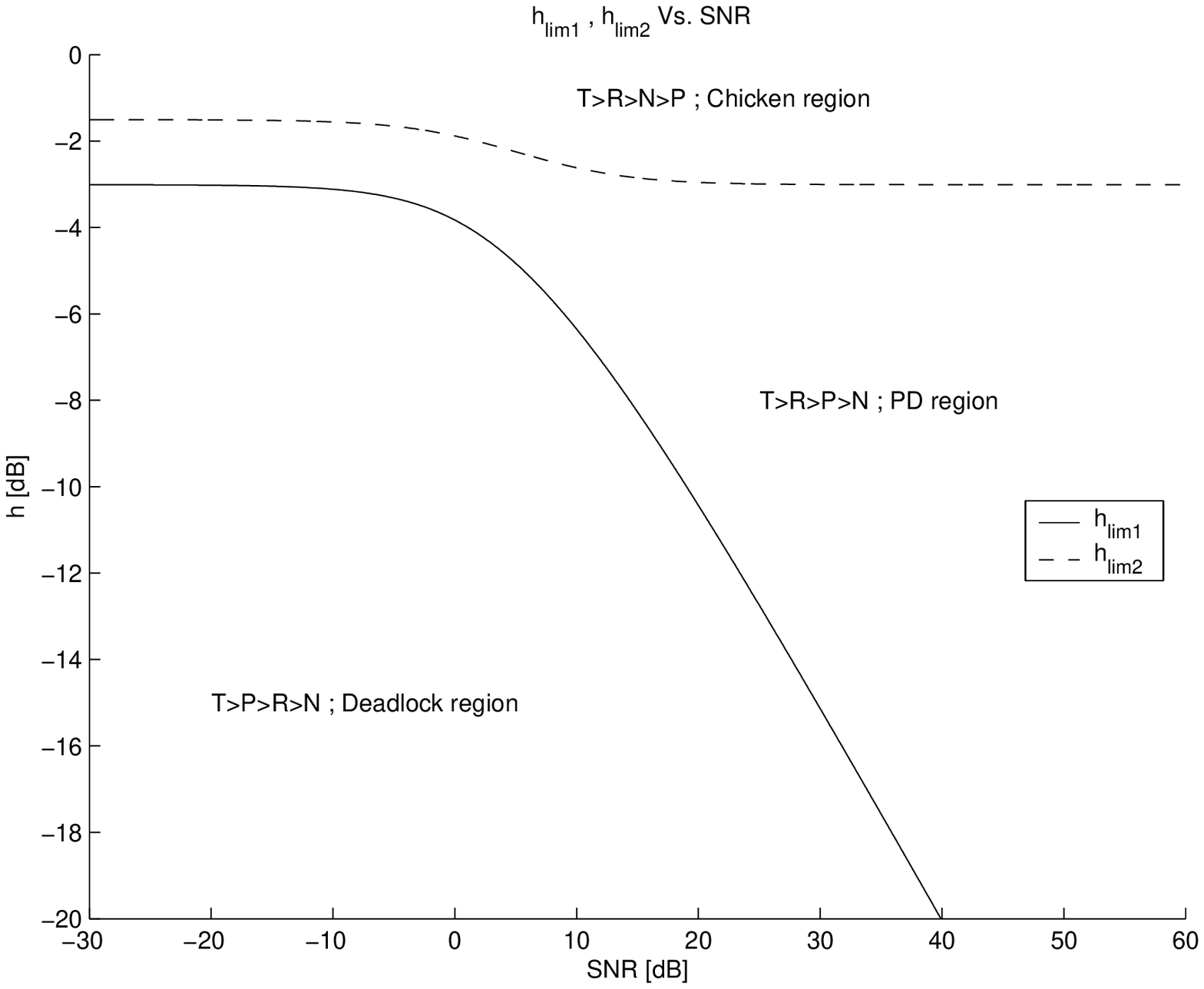}}
    \end{center}
    \caption{(a) Multiple Access Game with infinitely many pure strategy equilibria. Note that the parallel best response dynamics leads to unstable dynamics
   while the serial best response leads to one of two extremal NE points.
    (b) Graph of $h_{\lim1}$ , $h_{\lim2}$ Vs. SNR, The solid line
corresponds to $h_{\lim1}$ and the dashed line corresponds to $h_{\lim2}$ \label{fig 1} \cite{laufer2005}.}
  \label{pd1}
\end{figure}

 An important notion of solution relevant to games is that of a Nash equilibrium.
 \bdf
 A vector of actions $\va=(\va_1,...,\va_N)\in A$ is a Nash equilibrium
 in pure strategies if and only if for each player $1\le n \le N$ and for every $\va'=(a_1',...,a_N')$ such that $a'_i=\va_i$ for all $i \neq n$ and
 $a_n' \neq a_n$
 we have  $u_n(\va')<u_n(\va)$, i.e., each player can only loose by deviating by itself from the equilibrium.
 \edf
 The Nash equilibrium in pure strategies does not always exist as the following example shows:

 {\em Example I - A game with no pure strategy Nash equilibrium}: Consider the two players game defined by the following: $A_i=\{0,1\}$. $u_i(a_1,a_2)=a_1 \oplus a_2 \oplus (i-1)$, i.e., the first player
 payoff is $1$ when actions are identical and $0$ otherwise, while the second player's payoff is $1$ when the actions are different and $0$ otherwise.
 Clearly, this game also known as matching pennies has no Nash equilibrium in pure strategies, since always one of the players can improve his
 situation by changing his choice.  \\
 Even when it exists, the Nash equilibrium in pure strategies is not necessarily unique, as the following example shows:

 {\em Example II - A communication game with infinitely many pure strategy NE}. Assume that two users are sharing an AWGN
 multiple access channel (i.e., the accss point can perform joint decoding of the users)
 \beq
 y=x_1+x_2+z,
 \eeq
 where $z \sim N(0,\gs^2)$ is a Gaussian noise random variable. Each user has power $P$.  It is well known \cite{cover} that the rate region of this multiple access
 channel is given by a pentagon defined by:
\beq
\bea{l}
R_1 \le \frac{1}{2}\log \left(1+\frac{P}{\gs^2} \right)=C^{\max} \\
R_2 \le \frac{1}{2}\log \left(1+\frac{P}{\gs^2} \right)=C^{\max} \\
R_1+R_2 \le \frac{1}{2}\log \left(1+\frac{2P}{\gs^2} \right)=C_{1,2}. \\
\ena
\eeq
The corners $A,B$ (see figure \ref{mac_game}) of the pentagon are $A=\left(C^{\max},C^{\min}\right)$ and
$B=\left(C^{\min},C^{\max}\right)$, where $C^{\min}=\frac{1}{2}\log \left(1+\frac{P}{P+\gs^2} \right)$, is the rate achievable by assuming that the other
user's signal is interference. Note that any point on the line connecting the points $A,B$ is achieved by time
sharing between these two points. Each player $n=1,2$ can choose a strategy $0 \le \ga_n \le 1$ that is the time sharing ratio between coding at its
rate at point $A$ or $B$. The payoff in this game is given by
\beq
u_n(\ga_1,\ga_2)=\left \{
\bea{cc}
\ga_n C^{\max} + (1-\ga_n)C^{\min} & \hbox{if \ } \ga_1+\ga_2 \le 1 \\
0 & \hbox{otherwise.}
\ena
\right.
\eeq
The reason that the utility is $0$ when $\ga_1+\ga_2>1$ is that no reliable communication is possible, since the rate pair achieved is outside the
rate region. In this game any valid strategy point such that $\ga_1+\ga_2=1$ is a Nash equilibrium. If user $n$ reduces its $\ga_n$
obviously its rate is lower since he transmits larger fraction of the time at the lower rate. If on the other hand he increases $\ga_n$
then $\ga_1+\ga_2>1$ and both players achieve $0$. Hence the AWGN MAC game has infinitely many Nash equilibrium points.
Similar game has been used in \cite{gajic2008} where the fact that infinitely many NE points exist is shown.
It is interesting to note that a similar MAC game for the fading channel has a unique Nash equilibrium point \cite{lai2008}.

To better understand this game, we might look at the best response dynamics. Best response action is the attempt of a player to maximize its utility
against a given strategy vector. It is a well established mean of distributively achieving the Nash Equilibrium. In the context of information theory,
this strategy has been termed Iterative Water-Filling (IWF) \cite{yu2002}.
If in the multiple access game the players use the best response simultaneously, the first step would be to transmit at $C^{\max}$. Each player then
receives $0$ utility and in the next step reduces its rate to $C^{\min}$, and vice versa. The iteration never converges and the utility of each player
is given by $\frac{1}{2}C^{\min}$, worse than transmitting constantly at $C^{\min}$.  Interestingly in this case, the sequential best response leads
to one of the points $A,B$, which are the (non axis) corners of the rate region. The moral of this example is that using the best response strategy
should be done carefully even in multiple access scenario's such as in \cite{yu04}.
\subsection{Pure and mixed strategies}
To overcome the first problem of no equilibrium in pure strategies, the notion of mixed strategy has been proposed.
 \bdf
A mixed strategy $\pi_n$ for player $n$ is a probability distribution over $A_n$.
 \edf
 The interpretation of the mixed strategies is that player $n$ chooses his action randomly from $A_n$ according to the distribution $\pi_n$.
 The payoff of player $n$ in a game where the mixed strategies $\pi_1,...,\pi_N$ are played by the players is the expected value
 of the utility
 \beq
 u_n(\pi_1,...,\pi_N)=E_{\pi_1\times ... \times \pi_N} \left[u_n(x_1,...,x_N)\right].
 \eeq

{\em Example III: Mixed strategies in random access game over multiple access channel}
To demonstrate the notion of mixed strategy, we now extend the multiple access game, into a random multiple access game,
where the players can choose with probability $p_n$ of working at rate $C^{\min}$ and $1-p_n$ working at $C^{\max}$. This replaces the
synchronized TDMA strategy in the previous game, with slotted random access protocol.
This formulation, allows for two pure strategies corresponding to the corner points $A,B$ and the mixed strategies amount to
randomly choosing between these points. This game is a special case of the chicken dilemma (a termed proposed by B. Russel, \cite{poundstone92}),
since for each user it is better to chicken out, then to obtain zero rate, when both players choose the tough strategies.
Obviously from the previous discussion, the points $\left(C^{\max},C^{\min}\right)$ and $\left(C^{\min},C^{\max}\right)$ are Nash equilibria.
Simple computation shows that there is a unique Nash equilibrium in mixed strategy corresponding to $p_1=p_2=C^{\min}/C^{\max}$.
Interestingly the rates achieved by this random access (mixed strategy) approach is exactly
$\left(C^{\min},C^{\min}\right)$, i.e., the price paid for random access is that both players achieve their minimal rate, so simple
$p$-persistent random access provides no gain for the multiple access channel.
Following Papadimitriou \cite{papadimitriou2001} we can call this the price of random access.

\begin{table}
\centering
\caption{Payoffs in the multichannel random access game}
\begin{tabular}{c|c|c|}
I $\backslash$ II&  $0$  & $1$ \\
\hline
$0$ & $\left(C^{\min},C^{\min}\right)$ & $\left(C^{\min},C^{\max}\right)$ \\
\hline
$1$ & $\left(C^{\max},C^{\min}\right)$ & $\left(0,0\right)$ \\
\hline
\end{tabular}
\label{random_mac_game}
\end{table}

\subsection{convex games}
 An important special type of games which is important for the spectrum management problem is that of convex competitive games.
 \bdf
An $N$ player game $(N,A,U)$ is convex if each $A_n$ is compact and convex and each $u_n(x_1,...,x_n)$ is a convex function of $x_n$ for every choice
of $\{x_j:j \neq n\}$.
 \edf
 Convex games always have Nash equilibrium \cite{nikaido55}.
  A simple proof can be found in \cite{basar82}. Convex competitive games are especially important in the context of spectrum management,
  since the basic Gaussian interference game forms a convex game.
\subsection{The prisoner's dilemma}
 The prisoner's dilemma game is the major problem of the competitive approach. It was first described
 by Flood and Dresher in 1950 almost immediately after the concept of Nash equilibrium was published \cite{nash50}.
 For an overview of the prisoner's dilemma and its history see the excellent book of Poundstone \cite{poundstone92} or in this
 journal \cite{meshkati2007}.
 It turns out that this game has a unique Nash equilibrium which is the stable point of the game. Moreover this outcome
 is suboptimal to all player. The emergence of the prisoner's dilemma in simple symmetric interference channels has been discussed in
 \cite{laufer2005}. In \cite{leshem2006}, \cite{leshem2008} characterization of cases where cooperative solutions are better for general
 interference channels have been demonstrated. We briefly describe a simple case where the prisoner's dilemma occurs \cite{laufer2005}.
We assume a simplified two players game. The game is played over two frequency bands each with symmetric interference channel.
The channel matrices of this channel
are $\mH(1)=\mH(2)=\mH$ where
\[
|h_{12}(1)|^2=|h_{21}(1)|^2=|h_{12}(2)|^2=|h_{21}(2)|^2=h
\]
and $h_{ii}(k)=1$. We limit the discussion to $0\leq h < 1$.
In our symmetric game both users have the same power constraint $P$ and the power
is allocated by $p_1(1)=(1-\ga)P, p_1(2)=\ga P, p_2(1)=\gb P, p_2(2)=(1-\gb) P $. We assume that the decoder treats the interference as noise and cannot
decode it.
The utility for user I given power allocation parameters $\ga,\gb$ is given by its achievable rate
\begin{equation}
C^1=\frac{1}{2}\log_2\left(1+\frac{(1-\alpha)}{SNR^{-1}+\beta\cdot
h}\right)+\frac{1}{2}\log_2\left(1+\frac{\alpha}{SNR^{-1}+(1-\beta)\cdot
h}\right)
\end{equation}
and similarly for user II, we replace $\ga, \gb$.
The set of strategies in this simplified game is
$\left\{\alpha, \beta : 0\leq \alpha, \beta \leq 1\right\}$.
There are four alternatives outcomes:
\begin{itemize}
\item Both users select FDM resulting in $\alpha=\beta=0$. This is the reward (R) for cooperation.
\item Player I selects FDM while user II selects competitive approach (IWF) resulting in $\alpha=0$ ,
$\beta=(1-h)/2$ is the result of waterfilling by user II when $\alpha=0$). This is when player I is naive (N) and the temptation for player II (T)
\item Player I selects IWF while user II selects FDM  resulting in $\alpha=(1-h)/2$ , $\beta=0$.  This is the temptation for player I (T) and player II
is the naive (N).
\item Both players select IWF resulting in $\alpha=\beta=\frac{1}{2}$. This is the penalty for not cooperating (P).
\end{itemize}
Table \ref{payoff_table} describes the payoffs of users I at four different levels
of mutual cooperation (The payoffs of user II are the same with the
inversion of the cooperative/competetive roles).
\begin{table}[htbp]
\centering \caption{User I payoffs at different levels of mutual
cooperation}
\begin{tabular}{|c|c|c|}
\hline
&user II is fully cooperative&user II is fully competing\\
&$(\beta=0)$&$\left(\beta=\frac{(2\alpha-1)h+1}{2}\right)$\\
\hline $\begin{array}{c}
$user I is fully cooperative$ \\
(\alpha=0)
\end{array}$ & $\frac{1}{2}\log_2\left(1+\frac{1}{SNR^{-1}}\right)$&$\frac{1}{2}\log_2\left(1+\frac{1}{SNR^{-1}+\frac{(1-h)}{2}h}\right)$\\
\hline $\begin{array}{c}
$user I is fully competing$ \\
\left(\alpha=\frac{(2\beta-1)h+1}{2}\right)\\
\end{array}$ & $\frac{1}{2}\log_2\left(1+\frac{\frac{1+h}{2}}{SNR^{-1}}\right)+\frac{1}{2}\log_2\left(1+\frac{\frac{1-h}{2}}{SNR^{-1}+h}\right)$&$\log_2\left(1+\frac{\frac{1}{2}}{SNR^{-1}+\frac{1}{2}h}\right)$\\
\hline
\end{tabular}
\label{payoff_table}
\end{table}
A prisoner's dilemma situation is defined by the following payoff
relations for both players - $T>R>P>N$.
It is easy to show that the Nash equilibrium point in this case is
that both players will defect ($P$). This is caused by the fact that
given the other user act the best response will be to defect (since
$T>R$ and $P>N$). Obviously a better strategy (which makes this game
a dilemma) is mutual cooperation (since $R>P$).
We can now analyze this simple game.
It turns out that there are two functions, $h_{\lim_1}(SNR),h_{\lim_2}(SNR)$ as described in Figure \ref{pd_region} and only three
possible situations \cite{laufer2005}:
\begin{itemize}
\item (A)  $T>P>R>N$ , for $h<h_{\lim_1}$.
\item (B)  $T>R>P>N$ , for $h_{\lim_1}<h<h_{\lim_2}$.
\item (C)  $T>R>N>P$ , for $h_{\lim_2}<h$.
\end{itemize}
The payoff relations in (A) corresponds to a game called
"Deadlock". In this game there is no dilemma, since no matter what the other player does, it is better to
defect ($T>R$ and $P>N$), so the Nash equilibrium point is $P$.
Since $P>R$ thus there is no reason to
cooperate. The maximum sum rate is also $P$ because $2\cdot R>T+N$
and $P>R$. The payoff relations in (B) correspond to the prisoner's dilemma
situation. While the Nash equilibrium point is $P$, each user's maximum payoff is achieved by $R$.
In this region the FDM strategy will achieve the individual maximum rate.
The last payoff relations (C) corresponds to a game called
"Chicken". This game has two Nash
equilibrium points, $T$ and $N$. This is caused by the fact that for
each of the other player's strategies the opposite response is preferred
(if the other cooperates it is better to defect since $T>R$, while
if the other defects it is better to cooperate since $N>P$). The
maximum rate sum point is still at $R$ (since $R>P$ and $2\cdot
R>T+N$) thus, again FDM will achieve the maximum rate sum while IWF
will not.

\subsection{Generalized Nash games}
Games in strategic forms are very important part of game theory, and have many applications. However, in some cases the notion of a
game does not capture all the complications involved in the interaction between the players. Arrow and Debreu \cite{arrow54} defined
the concept of a generalized Nash game and generalized Nash equilibrium. In strategic form games, each player has a set of strategies that
is independent from the actions of the other players. However, in reality sometimes the actions of the players are constrained by the actions of the other
players. The generalized Nash game or abstract economy concept, captures exactly this dependence.
\bdf
A generalized Nash game with $N$ players, is defined as follows:
For each player $n$ we have a set of possible actions $X_n$ and a set function $K_n:\prod_{m \neq n} X_m \rarrow P(X_n)$ where $P(X)$
is the power set of $X$, i.e, $K_n\left(\left<x_m:m \neq n\right>\right) \subseteq X_n$ defines a subset of possible actions for player $n$
when other players play $\left<x_m:m \neq n\right>$. $u_n(\vx)$ is a utility function defined on all tuples $(x_1,...,x_N)$ such that
$x_n \in K_n(x_m: m \neq n)$.
\edf
Similarly to the definition of a Nash equilibrium, we can define a generalized Nash equilibrium:
\bdf
A generalized Nash equilibrium is a point $\vx=(x_1,...,x_N)$ such that for all $n$
$x_n \in K_n\left(\left<x_m: m \neq n\right>\right)$, and for all $\vy=(y_1,...,y_n)$ such that $y_n \in K_n(\left<x_m: m \neq n\right>)$ and $y_m=x_m$ for $m \neq n$
$u_n(\vx)\ge u_n(\vy)$.
\edf
Arrow and Debreu \cite{arrow54} proved the existence of a generalized Nash equilibrium
under very limiting conditions. Their result was generalized, and currently the best result is by Ichiishi \cite{ichiishi1983}.
This result can be used to analyze certain fixed rate and margin versions of the iterative water-filling algorithm \cite{pang_information_theory}
as will be shown in the next section.

\subsection{Nash bargaining theory}

The "Prisoner Dilema" highlights the drawbacks of competition, due to mutual mistrust of players. We
therefore ask ourselves when would a cooperative strategy
be preferable to a competitive strategy. It is apparent that the essential
condition for cooperation is that it should generate a
surplus, i.e. an extra gain which can be divided between the
parties. Bargaining is essentially the process of distributing the
surplus. Thus, bargaining is foremost, a process
in which parties seek to reconcile contradictory interests and
values. The main question is if all players commit to following the
rules, what reasonable outcome is acceptable by all parties.
Therefore, the players have to agree on a bargaining mechanism which
they will not abandon during the negotiations. The
bargaining results may be affected by several factors like the power
of each party, the amount of information available to each of the
players, the delay response of the players, etc.  Nash \cite{nash50}, \cite{nash53}
introduced an axiomatic approach that is based on properties that a
valid outcome of the bargaining should satisfy. This approach proved
to be very useful since it succeeded in choosing a unique solution
through a small number of simple conditions (axioms), thus saving
the need to perform the complicated bargaining process, once all
parties accept these conditions.

We now define the bargaining problem. An n-player bargaining
problem is described as a pair $\left<S,d \right>$, where $S$ is a convex set
in $n$-dimensional Euclidean space, consisting of all feasible sets
of $n$-players utilities that the players can receive when
cooperating. $d$ is an element of $S$, called the disagreement
point, representing the outcome if the players fail to
reach an agreement. $d$ can also can be viewed as the utilities
resulting from non-cooperative strategy (competition) between the
players, which is a Nash Equilibrium of a competitive game. The assumption
that $S$ is a convex set, is a reasonable assumption if both players
select cooperative strategies, since, the players can use alternating or mixed strategies to achieve convex combinations of pure
bargaining outcomes. Given a bargaining problem we say that the vector $\vu \in S$ is Pareto
optimal if for all $\vw\in S$ if $\vw\geq \vu$ (coordinate wise) then $\vw=\vu$.
A solution to the bargaining problem is a function $F$ defined on all bargaining problems such that
$F\left(\left< S,d \right> \right) \in S$.
Nash's axiomatic approach is based on the following four axioms that the solution function $F$ should satisfy:
\bds
\item[] {\em Linearity (LIN).} Assume that we consider the bargaining problem $\left<S',d'\right>$
obtained from the problem $\left<S,d \right>$ by transformations:
$s_n'=\ga_ns_n+\gb_n, \ \ n=1,...,N.$ $d_n'=\ga_nd_n+\gb_n$. Then
the solution satisfies $F_i\left(\left<S',d' \right>\right)=\ga_nF_n\left(\left<S,d \right>\right)+\gb_n$, for all
$n=1,...,N$.

\item[]{\em Symmetry (SYM)}. If two players $m<n$ are identical in the sense that $S$ is symmetric with respect to
changing the $m$'th and the $n$'th coordinates, then $F_m\left(\left<S,d \right>\right)=F_n\left(\left<S,d \right>\right)$. Equivalently,
players which have identical bargaining preferences, get the same
outcome at the end of the bargaining process.
\item[]{\em Pareto optimality (PAR)}. If $\vs$ is the outcome of the bargaining then no other
state $\vt$ exists such that $\vs<\vt$ (coordinate wise).
\item[]{\em Independence of irrelevant alternatives (IIA)}. If $S\subseteq
T$ and if $F(\left<T,d \right>)=\left(u_1^*, u_2^*\right)\in S$, then
$F(\left<S,d \right>)=\left(u_1^*, u_2^*\right)$. \eds
%
%

Surprisingly these simply axioms fully
characterize a bargaining solution titled Nash's Bargaining solution. Nash's theorem may be stated as follows \cite{nash50}.

Based on this axioms and definitions we now can state Nash's
theorem.
\bthm \label{Nash's_theorem}
Assume that for all $S$ is
compact and convex, then there is a unique bargaining solution
$F\left(\left<S,d\right>\right)$, satisfying the axioms INV, SYM, IIA, and PAR, which is
given by
\beq
F\left(\left<S,d\right>\right) = (s_1,..,s_N)=\arg \max_{(d_1,..,d_N)\leq
(s_1,..,s_N)\in S} \prod_{n=1}^N \left(s_n-d_n\right).
\eeq
\ethm

Before continuing the study of the bargaining solution, we add a cautionary word. While, Nash's axioms are mathematically appealing, they may not be
acceptable in some scenarios.
Indeed various alternatives to these axioms have been proposed that lead to other solution concepts.
Extensive survey of these solutions can be found in \cite{kalai556}. In the communication context, the axioms proposed by
Boche at. el.  \cite{boche07} lead to a generalized NBS solution. More results of generalized bargaining for frequency selective channels
will be discussed in \cite{zehavi2009}.
To demonstrate the use of the Nash bargaining solution to interference channels, we begin with a simple example
for flat channels.

{\em Example III}: Consider two players communicating over a 2x2
memoryless Gaussian interference channel with bandwidth $W=1$, as described in (\ref{standard_IC}.
Assume for simplicity that $P_1=P_2=P$. We assume that no receiver can perform joint
decoding, and the utility of player $n$, $U_n$, is given by the
achievable rate $R_n$. Similarly to the prisoner's dilemma example, if the players choose to compete then the
competitive strategies in the interference game are given by flat
power allocation and the resulting rates are  given by $R_{nC}=1/2 \log_2 \left(1+\frac{P}{1+\ga_n^2 P}\right)$.
Since the rates $R_{nC}$ are achieved by competitive strategy, player
$n$ would  cooperate only if  he will obtain a rate higher than
$R_{nC}$. The game theoretic rate region is defined by pair rates
higher than the competitive rates $R_{nC}$ (see figure \ref{game_region}). Assume that
the players agree on using only FDM cooperative strategies. i.e.
player $n$ uses a fraction of
 $0 \le \gr_n \le 1$ of the band. If we consider only Pareto optimal strategy vectors, then obviously $\gr_2=1-\gr_1$.
  The rates obtained by the two users are given by
 $R_n(\gr_n)=\frac{\gr_n}{2} \log_2 \left(1+\frac{P}{\gr_n} \right)$.
 The two users will benefit from FDM type of cooperation as long as
 \beq
 R_{nC} \le R_n(\gr_1), \ \ \ n=1,2.
 \eeq
The FDM Nash bargaining solution is given by solving the problem
 \beq
 \max_\gr F(\gr)=\max_{\gr}\left(R_1(\gr)-R_{1C} \right) \left(R_2(\gr)-R_{2C} \right).
 \eeq
The cooperative solution for this flat channel model was derived in
\cite{leshem2006}, \cite{leshem2008}.

\section{Application of game theory to frequency selective interference channels}
\label{game_theory_concepts}
In this section we apply the ideas presented in previous sections to analyze the frequency selective interference game.
We provide examples for both competitive and cooperative game theoretic concepts.
\subsection{Waterfilling based solutions and the Nash equilibrium}
\label{sec:GI_game} To analyze the competitive approach to frequency selective interference channels, we first
define the discrete-frequency Gaussian
interference game which is a discrete version of the game defined in \cite{yu2002}.
Let $f_0 < \cdots <f_K$ be an increasing sequence of
frequencies. Let $I_k$ be the closed interval given by
$I_k=[f_{k-1},f_k]$. We now define the approximate Gaussian
interference game denoted by $GI_{\{I_1, \ldots, I_K\}}$.

Let the players $1,\ldots,N$ operate over $K$ parallel channels. Assume
that the $K$ channels have coupling functions $h_{ij}(k)$.
Assume that user $i$ is allowed to transmit a total power of
$P_i$. Each player can transmit a power vector $\vp_n=\left(
p_n(1),\ldots,p_n(K) \right)  \in [0,P_n]^K$ such that $p_n(k)$ is
the power transmitted in the interval $I_k$. Therefore we have $\sum_{k=1}^K p_n(k)=P_n$.
The equality follows from the fact that in
a non-cooperative scenario all users will use all the available power.
This implies that the set of power distributions for all
users is a closed convex subset of the hypercube $\prod_{n=1}^N
[0,P_n]^K$ given by: \beq \label{eq_strategies} \mB=\prod_{n=1}^N
\mB_n \eeq where $\mB_n$ is the set of admissible power
distributions for player $n$ given by:
\beq
\mB_n=[0,P_n]^K\cap \left\{\left(p(1),\ldots,p(K)\right): \sum_{k=1}^K p(k)=P_n \right\}
\eeq
Each player chooses a PSD $\vp_n=\left<p_n(k): 1\le k \le N
\right > \in \mB_n$. Let the payoff for user $i$ be given by: \beq
\label{eq_capacity}
C_n\left(\vp_1,\ldots,\vp_N\right)= \\
\sum_{k=1}^{K}\log_2\left(1+\frac{|h_n(k)|^2p_n(k)}{\sum
|h_{nm}(k)|^2 p_m(k)+\gs^2_n(k)}\right)
\eeq where $C_n$ is the capacity available to player $n$ given power
distributions $\vp_1,\ldots,\vp_N$, channel responses $h_n(f)$,
crosstalk coupling functions $h_{mn}(k)$ and $\gs^2_n(k)>0$ is external
noise present at the $i$'th channel receiver at frequency $k$. In
cases where  $\gs^2_n(k)=0$ capacities might become infinite using FDM
strategies, however this is non-physical situation due to the
receiver noise that is always present, even if small. Each $C_n$ is
continuous in all variables.
\begin{definition}
The Gaussian Interference game $GI_{\{I_1,\ldots,I_k\}}=\left\{N,\mB,\mC\right\}$ is the N
players non-cooperative game with payoff vector
$\mC=\left(C_1,\ldots,C_N \right)$ where $C_n$ are defined in
(\ref{eq_capacity}) and $\mB$ is the strategy set defined by (\ref{eq_strategies}).
\end{definition}

The interference game is a convex non-cooperative N-persons
game, since each $\mB_i$ is compact and convex and each $C_n(\vp_1,...,\vp_N)$ is continuous and convex in $\vp_n$ for any value of $\{\vp_m, m\neq n\}$.
Therefore it always has a Nash equilibrium in pure strategies. A presentation of the proof in this case using waterfilling interpretation is
given in \cite{laufer2006}.

A much harder problem is the uniqueness of Nash equilibrium points in
the water-filling game. This is very important to the stability of the
waterfilling strategies. A first result in this direction has been
given in \cite{chung2002}. A more general analysis of the convergence
has been given in  \cite{yu2002},
\cite{Chung2003},\cite{Luo2005},\cite{scutari2006} and \cite{Shum2007}.
Even the uniqueness of the Nash equilibrium, does not imply a stable dynamics. Scutari et al. \cite{scutari2008}
provided conditions for convergence. Basically, they use the Banach fixed point theorem, and require that the waterfilling response will
be a contractive mapping. The waterfilling process has several versions: The sequential IWF is performed by a single player at each step. The
parallel IWF is performed simultaneously by all players at each step, and the asynchronous IWF is performed in an arbitrary order. For good discussion
of the convergence of these techniques see \cite{scutari2008}. It should be emphasized that some conditions on the interference channel
matrices are indeed required. A simple condition is strong diagonal dominance \cite{yu2002}, and other papers relaxed these assumptions.
In all typical DSL scenarios, the IWF algorithms converge. However, the convergence conditions are not always met,
even in very simple cases, as the following examples shows:

{\em Example IV - Divergence of the parallel IWF.} We consider a Gaussian interference game with 2 tones and 3 players. Each player has total power $P$.
The signal received by each receiver is just $y_n(k)=\sum_{m=1}^3 x_m(k) + z_n(k)$ where
$z_n(k)~N\left(0,\gs^2(k)\right)$, where the noise in the second band is stronger satisfying $\gs^2(2)=P+\gs^2(1)$.
We assume simultaneous waterfilling is performed (Similar examples can be constructed for the sequential IWF algorithm, but they are
omitted due to space limitations). At the first stage, all users put all their power into frequency $1$, by the first inequality.
At the second stage all users see noise and interference power of $2P+\gs^2(1)$ at the first frequency, while the interference at the
second frequency is $\gs^2(2)=P+\gs^2(1)$. Since even when all power is put into frequency $2$ the total power + noise is below the noise level
at frequency $1$ all users will move their total power to frequency $2$. This will continue, with all users alternating between frequenies
simultaneously. The average rate obtained by the simultaneous iterative waterfilling is
\[
\frac{1}{4} \log\left(1+\frac{P}{2P+\gs^2(1)} \right)+\frac{1}{4} \log\left(1+\frac{P}{3P+\gs^2(1)} \right)
\]
Nash equilibrium exists in this case, for example, two users use frequency $1$ and one user uses frequency $2$ resulting in a rate
\[
\frac{1}{2}\log\left(1+\frac{P}{P+\gs^2(1)} \right)
\]
for each user.

The previous example demonstrated a simple condition where one of the water-filling schemes diverges. However, there are multiple NE points.
The situation can be even worse. The following channel is a frequency selective channel, with a single NE in the Gaussian interference game. However
non of the water filling schemes converges.

\begin{figure}
    \begin{center}
    \subfigure[Channel where the IWF fails to converge]
    {
    \label{triangle_interferene}
    \includegraphics[width=0.35\textwidth]{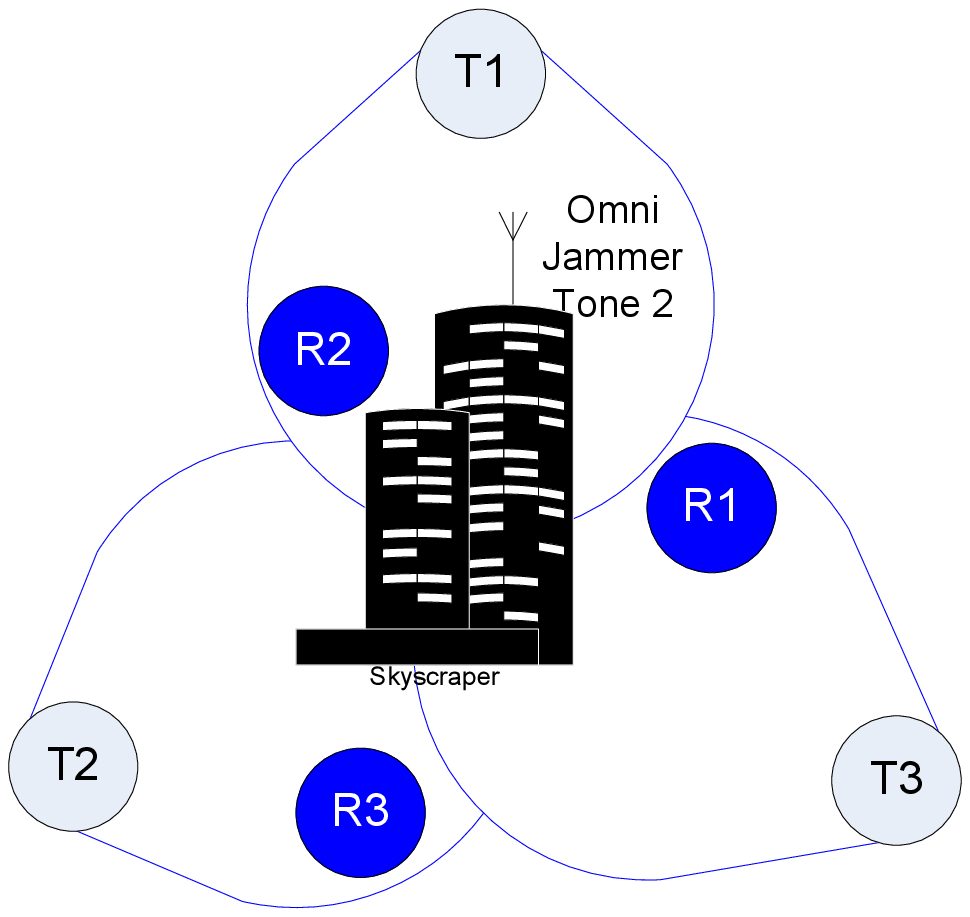}}
    \subfigure[Game theoretic rate region]
    {
    \label{game_region}
    \includegraphics[width=0.4\textwidth]{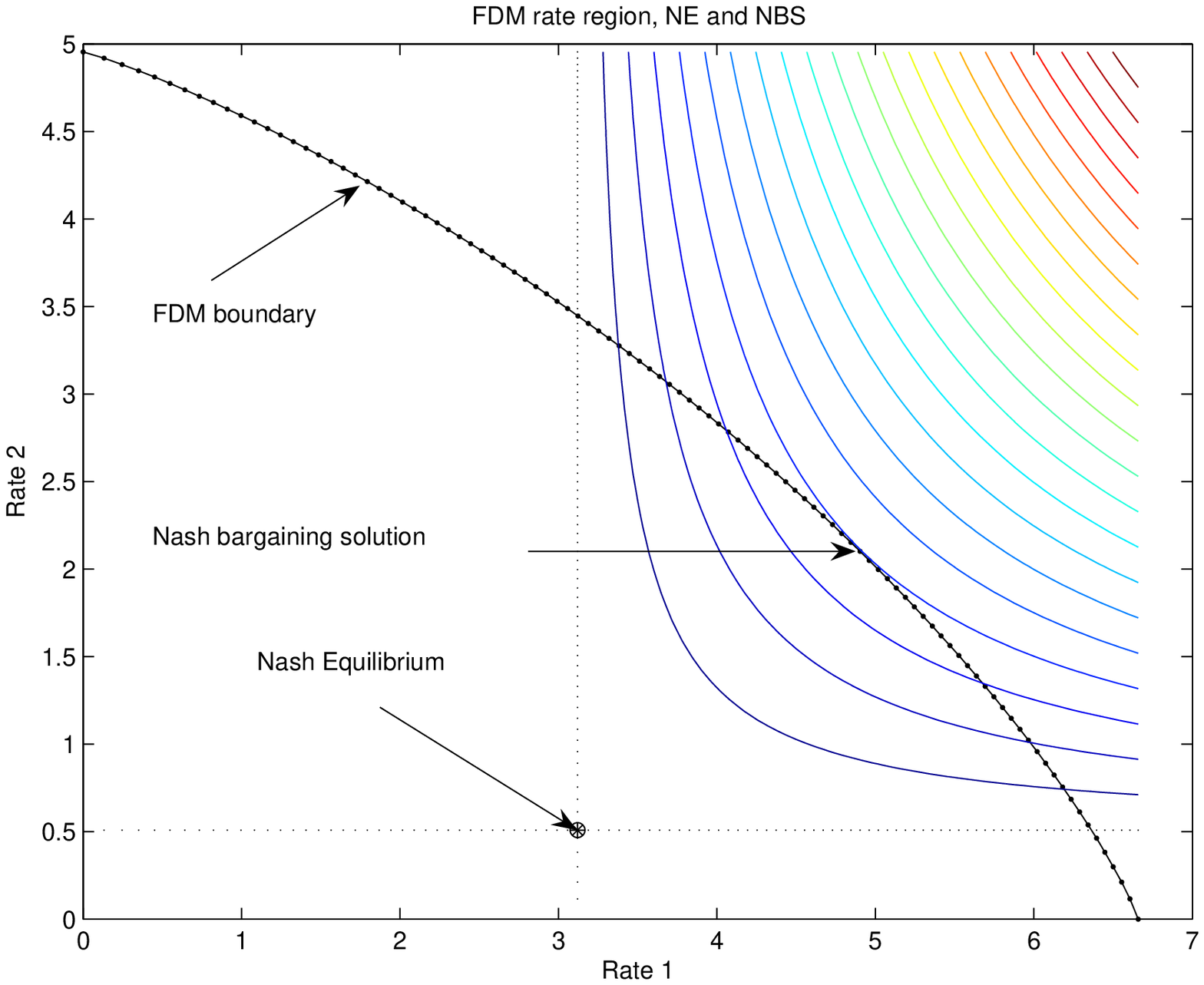}}
    \end{center}
    \caption{ (a) A frequency selective channel with unique NE, where both sequential and parallel IWF diverge. (b)The Game Theoretic Rate region is defined by the set of rates which are better than the competitive equilibrium. The boundary of the region is
    exactly the set of Pareto-optimal points. The Hyperbola tangent to the rate region defines the Nash bargaining solution.}
  \label{regions}
\end{figure}

{\em Example V - Divergence of all waterfilling approaches.} We provide now a second example, where both the sequential and the parallel IWF diverge,
even though there is a unique NE point. Assume that we have two channels where
the channel matrices $\mH(k), k=0,1$ are equal and given by:
\beq
\mH(k)=\left[
\bea{ccc}
1 & 0 & 2 \\
2 & 1 & 0 \\
0 & 2 & 1
\ena
\right]
\eeq
and the noise at the first tone is $\gs^2$ and at the second tone $\gs^2+P$.
Each user has total power $P$. This situation might occur when there is a strong interferer at tone $2$ while the receivers and transmitters are
located on the sides of a triangle, each user transmitting to a receiver near the next transmitter as in figure \ref{triangle_interferene}.
When the first user allocates its power it puts all its power at the first frequency. The second user chooses tone 2. The third player puts all its power
at frequency $1$, but this generates interference to user $1$ which migrates to frequency $2$ and we obtain that the users change their transmission
tone at each step. In the simultaneous IWF all users will choose channel 1 and then migrate together to channel 2 and back. It is interesting that
this game has a unique equilibrium, where each user allocates two thirds of the power to frequency 1 and one third of the power to frequency 2. Still all
iterative schemes diverge.

%
\subsection{Pricing mechanisms for regulating distributed solutions}
To overcome some of the problems of the competitive behavior, regulation can play an important role.
A generalization of the RA-IWF algorithm is the band preference (BPSM) algorithm \cite{Cioffi_BPSM} in which each user
solves the following problem in parallel (or sequentially) to the other users:
\beq
\label{BPSM}
\bea{ll}
\max_{p_n(1),\ldots,p_n(K)} & \sum_{k=1}^K c_n(k)\log_2\left(1+\frac{|h_{nn}(k)|^2
p_n(k)}
{\sum_{m \neq n} |h_{nm}(k)|^2 p_m(k)+\gs^2_n(k)}\right) \\
\hbox{Subject to } &  P_n=\sum_{k=1}^K p_n(k) \;.
\ena
\eeq
In the BPSM algorithm the total rate is replaced by a weighted sum of the rates at each frequency. The weights can be provided by the regulator
to limit the use of certain bands by strong users, so that users that suffers severe interference but do not affect other users will be protected.
This results in waterfilling against a compensated noise level.

Alternative approach to the BPSM is using generalized Nash games. The basic approach has been proposed in \cite{yu2002} and termed Fixed Margin IWF.
Each user has both power constraint, a target noise margin and a desired rate. The user minimizes its power as long as it achieves its target rate.
This is a generalized Nash game, first analyzed by Pang et al. \cite{pang_information_theory}, who provided the first conditions for convergence.
Noam and Leshem \cite{noam2009} proposed a generalization of the FM-IWF termed Iterative Power Pricing (IPP).
In their solution, a weighted power is minimized, where frequency bands which have higher
capacity are "reserved" to players with longer lines, through a line dependent pricing mechanism. The users iteratively optimize their
power allocation so as their rate and total power constraint are maintained while minimizing the total weighted power. It can be shown that the
conditions of Pang et al. can also be used to analyze the IPP algorithm. For both the BPSM and the IPP approaches, simple pricing schemes that are adapted to
the DSL scenario have been proposed. The general question of finding good pricing schemes is still open, but would require to combine physical
modeling of the channels as well as insight into the utilities as function of the desired rate. Even the autonomous spectrum balancing algorithm
(ASB) \cite{cendrillon2007} can be viewed as a generalized Nash game, where the utility is given by the rate of a fictitious reference line, and the
strategy sets should satisfy both rate and power constraint. The drawback of the ASB approach here, is finding a reference line which serves as
a good utility function.
\subsection{Bargaining over frequency selective channels under mask
constraint } \label{freq_selective}

In this section we define the cooperative game corresponding to the
joint FDM/TDM achievable rate region for the frequency selective $N-$
user interference channel. For simplicity of presentation  we limit
the derivation to the two player case under PSD mask constraint.
In \cite{han2005} Nash bargaining solution was used for resource allocation in OFDMA systems. The
goal was to maximize the overall system rate, under constraints on
each user's minimal rate requirement and total transmitted power.
However, in that paper the solution was used only
as a measure of fairness. Therefore, the point of disagreement was
not taken as the Nash equilibrium for the competitive game, but an
arbitrary $R_{min}$ was used. This can result in a non-feasible problem, and
the proposed algorithm might become unstable.
An alternative approach is based on PSD mask constraint \cite{leshem2008}
in conjunction with general bargaining theory originally developed by Nash (\cite{nash50},\cite{nash53}).
Based on the solution for the PSD limited case, computing the Nash Bargaining solution under total power constraint can then be solved
efficiently as well \cite{zehavi2009}. In order to keep the discussion simple we concentrate the discussion on the two user PSD mask limited case.

In real applications, the regulator limits the PSD mask and not only
the total power constraint. Let the $K$ channel matrices at
frequencies $k=1,...,K$ be given by $\left<\mH_k:k=1,...,K\right>$.
Player is allowed to transmit at maximum power
$p_n\left(k\right)$ in the $k$'th frequency bin. In competitive
scenario, under mask constraint, all players transmit at the maximal
power they can use. Thus, player $n$ choose the PSD,
$\vp_n=\left<p_n(k): 1\le k \le K \right >$. The payoff for user $n$
in the non-cooperative game is therefore given by:
\begin{equation}
\label{eq_capacityA}
 R_{nC}\left(\vp\right)=
 \sum_{k=1}^{K}\log_2\left(1+\frac{|h_{nn}(k)|^2p_n(k)}{\sum_{m \neq n}
 |h_{nm}(k)|^2 p_m(k)+\gs_n^2(k)}\right).
 \end{equation}
 Here, $R_{nC}$ is the capacity available to player $n$ given a PSD mask constraint
 distributions $\vp$.  $\gs_n^2(k)>0$ is the noise presents at the $n$'th receiver at
 frequency $k$. Note that without loss of generality, and in order to simplify notation,
we assume that the width of each bin is normalized to 1.
 We now define the cooperative game $G_{TF}(2,K,\vp)$.
\begin{definition}
The FDM/TDM game $G_{TF}(2,K,\vp)$ is a game between $2$ players
transmitting over $K$ frequency bins under common PSD mask
constraint. Each player has full knowledge of the channel matrices
$\mH_k$ and the following conditions hold:
\begin{enumerate}
\item Player $n$ transmits using a PSD limited by $\left<p_n(k): \ k=1,...,K \right>$.
\item The players are using coordinated FDM/TDM strategies. The strategy for player $1$ is a  vector
$\vga_1=[\ga_1(1),...,\ga_1(K)]^T$ where $\ga_1(k)$ is the proportion of
time player $1$ uses the $k$'th frequency channel. Similarly, the
strategy for player $2$ is a  vector
$\vga_2=[\ga_2(1),...,\ga_2(K)]^T$.
\item The utility of the players is given by
\begin{equation}
R_n\left(\vga_n\right)
   =\sum_{k=1}^K \alpha_n\left(k\right)R_n(k)
        = \sum_{k=1}^K \ga_n(k)\log_2
\left(1+\frac{|h_{11}(k)|^2 p_n(k)}{\gs_n^2(k)}\right). \\
 \end{equation}
\end{enumerate}
\end{definition}
By Pareto optimality of the Nash Bargaining solution for each $k$, $\ga_2(k)=1-\ga_1(k)$,
so we will only refer to $\vga=\vga_1$ as the strategy for both players.
Note, that interference is avoided by time sharing at
each frequency band, i.e only one player transmits with maximal
power at a given frequency bin at any time. The allocation of the
spectrum using the vector $\vga$ induces a simple
convex optimization problem that can be posed as follows
\begin{equation}
\begin{array}{c}
\textbf{max}
\left(R_1(\boldsymbol{\alpha})-R_{1C}\right)\left(R_2(\boldsymbol{\alpha})-R_{2C}\right)\\
\textbf{subject to: } 0\leq\alpha(k)\leq1  \quad \forall k,
R_{nC}\leq R_n\left(\boldsymbol{\alpha}\right) \quad \forall n.
\end{array}
\label{nash_FDM}
\end{equation}
since the $\log$ of the Nash function (\ref{nash_FDM}) is a convex function the overall problem is convex. Hence, it
can be solved efficiently using KKT conditions \cite{leshem2008}.
Assuming that a feasible
solution exists it follows from the KKT conditions that the allocation is done according to the
following rules:
\begin{enumerate}
\item  The two players are sharing the frequency
bin $k$, ($0<\ga(k)<1$) if
\begin{equation}
\label{rule1}
 \frac{R_{1}\left(k\right)}
     {R_1\left(\boldsymbol{\alpha}\right)-R_{1C}}=
     \frac{R_{2}\left(k\right)}{R_2\left(\boldsymbol{\alpha}\right)-R_{2C}}.
 \end{equation}
\item Only player $n$ is using the frequency bin $k$, ($\alpha_n\left(k\right)=1$),
if
\begin{equation}
 \frac{R_{n}\left(k\right)}{R_n\left(\boldsymbol{\alpha}\right)-R_{nC}}>\frac{R_{3-n}\left(k\right)}{R_{3-n}\left(\boldsymbol{\alpha}\right)-R_{3-n,C}}.
 \end{equation}
\end{enumerate}

These rules can be further simplified. Let
$L_{k}=R_1(k)/R_2(k)$ be the ratio between the rates at each frequency
bin. We can sort the frequency bins in decreasing  order
according to $L_{k}$. {\em From now on we assume that when $k_1<k_2$ then $L_{k_1}>L_{k_2}$}. If all the values of $L_{k}$ are distinct then
there is at most a  single frequency bin that has to be shared
between the two players. Since only one bin can satisfy
equation (\ref{rule1}), let us denote this frequency bin as $k_s$,
then all the frequency bins $1\leq k<k_s$ will only be used by
player $1$, while all the frequency bins $k_s < k \leq K$ will be
used by player $2$. The frequency bin $k_s$ has to be
shared according to the rules.

We now have to find the frequency bin that has to be shared between
the players if there is a solution. Let us define the surplus of
players $1$ and $2$ when using Nash bargaining solution as
$A=\sum_{m=1}^K\alpha\left(m\right)R_1\left(m\right)-R_{1C}$,
 and $B=\sum_{m=1}^K\left(1-\alpha\left(m\right)\right)R_2\left(m\right)-R_{2C}$,
respectively. Note that the ratio, $\Gamma=A/B$ is independent of frequency and is set by the optimal
assignment. A-priori  $\Gamma$ is unknown and may not exists. We are
now ready to define the optimal assignment of the $\alpha$'s.

Let $\Gamma_k$ be a moving threshold defined by $\Gamma_k=A_k/B_k.$
where \beq A_k=\sum_{m=1}^k R_1\left(m\right)-R_{1C},
B_k=\sum_{m=k+1}^K R_2\left(m\right)-R_{2C}.\eeq $A_k$ is a
monotonically increasing sequence, while $B_k$ is monotonically
decreasing. Hence, $\Gamma_k$ is also monotonically increasing.
$A_k$ is the surplus of player $1$  when frequencies
$1,...,k$ are allocated to player $1$. Similarly $B_k$ is the
surplus of player $2$ when frequencies $k+1,...,K$ are allocated to
player $2$.

Let $k_{\min}=\min_k\left\{k: A_k \ge 0\right\},$ and
$k_{\max}=\min_k\left\{k:B_k < 0\right\}$.

Since we are interested in feasible NBS, we must have positive
surplus for both users. Therefore, using the allocation rules, we obtain
$k_{\min}\le k_{\max}$ and $ L_{k_{\min}} \le \Gamma \le
L_{k_{\max}}$. The sequence $\{\Gamma_m: k_{min} \le m \le
k_{max}-1\}$ is strictly increasing, and always positive.
While the threshold $\Gamma$ is unknown, one can use the sequences
$\Gamma_k$ and $L(k)$ to find the correct $\Gamma$.

If there is a Nash bargaining solution, let $k_s$ be the frequency
bin that is shared by the players. Then, $k_{\min}\leq k_s\leq
k_{\max}$. Since, both players must have a positive gain in the game
($A>A_{k_{\min}-1}$,$B>B_{k_{\max}}$). Let $k_s$ be the smallest
integer such that $L(k_s)<\Gamma_{k_s}$, if such $k_s$ exists.
Otherwise let $k_s=k_{\max}$.
 \blm{} The following two statements provide the solution
 \label{lemmax}
\begin{description}
\item [1] If a Nash bargaining solution exists for $k_{\min}\leq k_s < k_{\max}$,
then $\alpha\left(k_s\right)$ is given by
$\alpha\left(k_s\right)=max\{0, g\},$ where \beq g=
1+\frac{B_{k_s}}{2 R_2\left(k_s\right)}\left(1-
\frac{\Gamma_{k_s}}{L(k_s)}\right). \eeq
\item [2] If a Nash bargaining solution exists and there is no such $k_s$, then
$k_s=k_{max}$ and  $\alpha\left(k_s\right)=g$.
\end{description}
\elm
Based on the pervious lemmas the algorithm is described in table
\ref{two_players_table}. In the first stage, the algorithm computes
$L(k)$ and sorts them in a non increasing order. Then $k_{\min},
k_{\max}, A_k,$ and $B_k$ are computed. In the second stage the
algorithm computes  $k_s$ and $\vga$.

\begin{table}
\centering \caption{Algorithm for computing the 2x2 frequency
selective NBS}
\begin{tabular}{||l||}
\hline \hline {\bf: Initialization:}  Sort the ratios $L(k)$
in decreasing order. \\
Calculate the values of $A_k,B_k$ and $\Gamma_k, k_{\min}, k_{\max}$, \\
\hline
If $k_{\min}>k_{\max}$ no NBS exists. Use competitive solution. \\
Else \\
\quad For $k=k_{\min}$ to $k_{max}-1$ \\
 \quad \quad if $L(k)\leq\Gamma_k$. \\
\quad \quad \quad Set $k_s=k$ and  $\alpha'$s according to the
lemmas-This is NBS. Stop\\
\quad \quad End \\
\quad End \\
\quad If no such $k$ exists, set $k_s=k_{\max}$ and calculate $g$. \\
\quad If $g\geq0$ set $\alpha_{k_s}=g, \alpha(k)=1$, for $k<k_{\max}$. Stop. \\
\quad Else ($g<0$) \\
 \quad \quad There is no NBS. Use competitive solution. \\
 \quad End. \\
End \\
\hline \hline
\end{tabular}
\label{two_players_table}
\end{table}
To demonstrate the algorithm we compute the Nash bargaining for the following example:

{\em Example V}: Consider two players communicating over 2x2
memoryless Gaussian interference channel with 6 frequency bins. The
players' rates if they are not cooperate is $R_{1C}=15$, and
$R_{2C}=10$. The feasible rates $R_1\left(k\right)$ and
$R_2\left(k\right)$  in each frequency bin with no interference are
given in Table \ref{sorttable} after sorting the frequency bins with
respect to $L_k$.
\begin{table}
\centering
\begin{tabular}{|c|c|c|c|c|c|c|}

  \hline
 $k$                 & 1 & 2 & 3 & 4  & 5 & 6  \\
 \hline
 $R_1$            & 14 & 18 & 5 & 10 & 9 & 3 \\
  \hline
 $R_2$            & 6 & 10 & 5 &15 & 19&19 \\
  \hline
 $L\left(k\right)$ &2.33 &1.80 &1.00& 0.67& 0.47 &0.16 \\
  \hline
 $A_k$ & -1 & 17 & 22 &32 &41 &44 \\
   \hline
 $B_k$ & 58 & 48& 43& 28& 9 &-10 \\
 \hline
 $\Gamma_{k}$ & -0.02& 0.35 &0.51 &1.14 &4.56 &-4.40 \\
\hline
\end{tabular}

  \caption{The rates of the players in each frequency bin after sorting.}
  \label{sorttable}
\end{table}

We now have to compute the surplus $A_k$ and $B_k$  for each
players. If NBS exists then the players must have postive surplus,
thus, $k_{min}=2$, and $k_{max}=4$. Since, $k=4$ is the first bin
such that  $\Gamma_{k}> L_k$, we can conclude that $k_s=4$, and
$\alpha=0.33$ (using lemma \ref{lemmax}) . Thus, players $1$ is
using frequency bins $1,2$, and $3$, and using  $1/3$ of the time
frequency  bin $4$. The total rate of players $1$ and $2$ are $40\frac{1}{3}$
and $48$ respectively.

\begin{figure}
    \begin{center}
    \subfigure[The FDM/TDM rate region]
    {
    \label{FDM_rate_region}
    \includegraphics[width=0.35\textwidth]{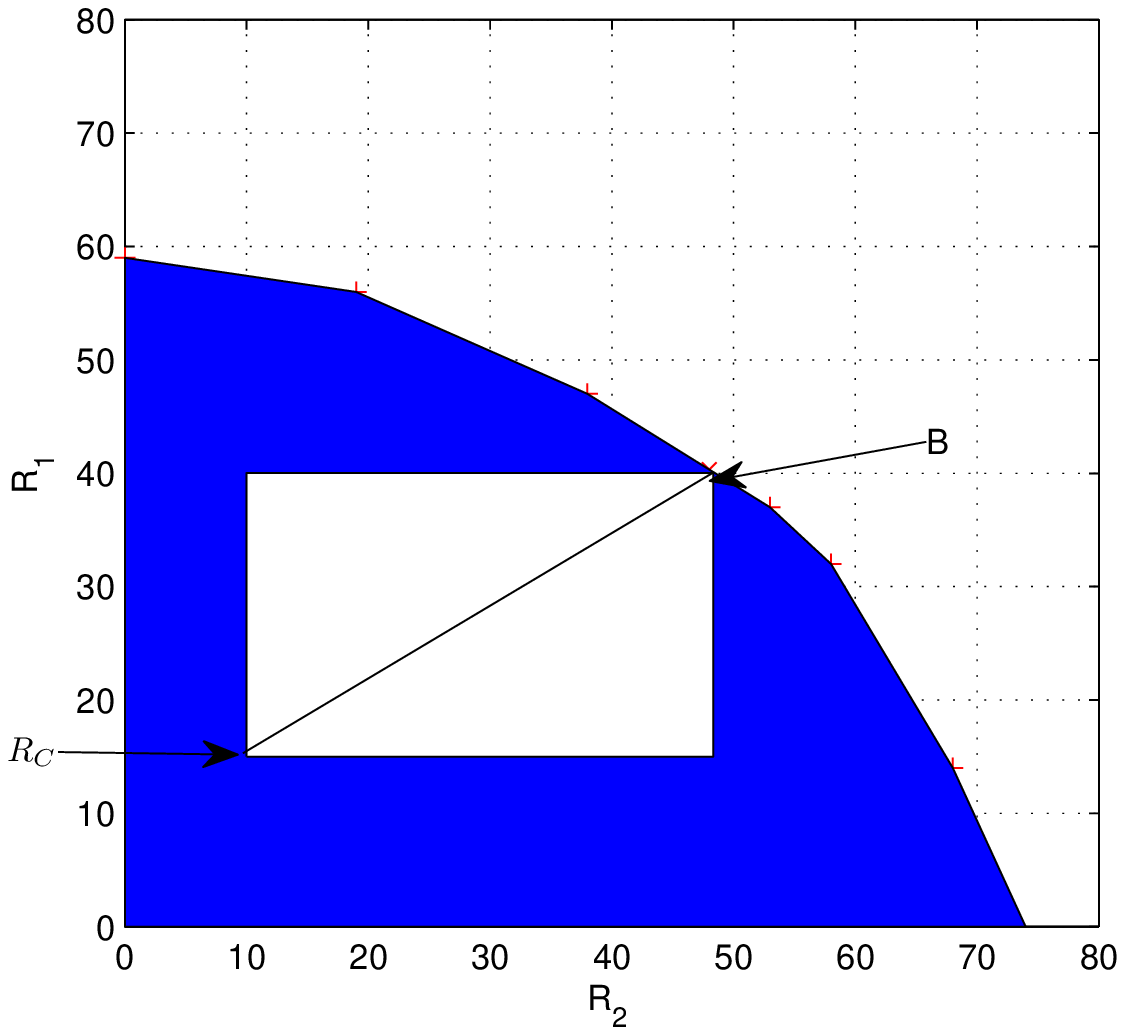}}
    \subfigure[Price of Anarchy for 32 parallel Rayleigh fading channels]
    {
    \label{NBS}
    \includegraphics[width=0.35\textwidth]{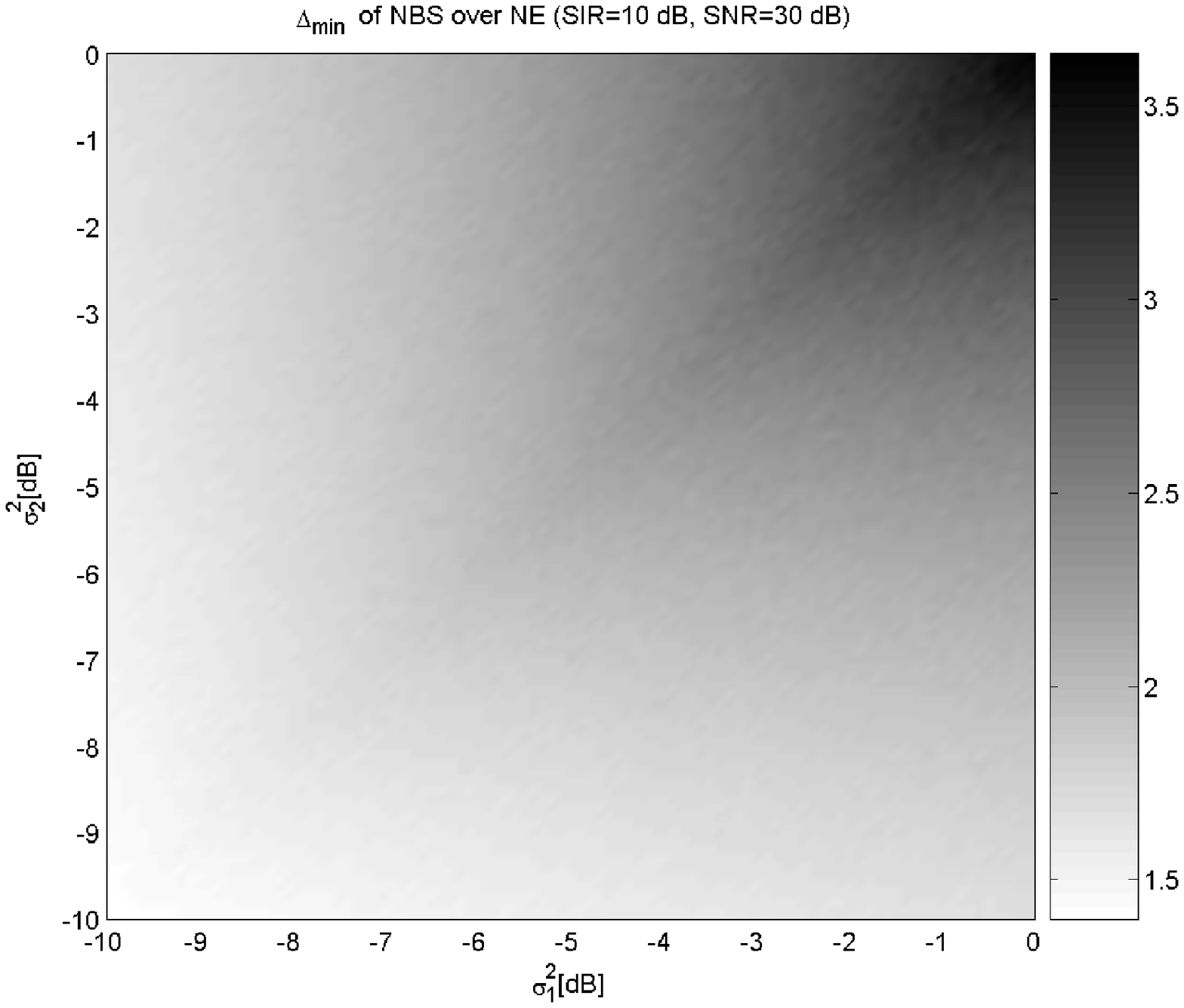}}
    \end{center}
    \caption{(a) Feasible FDM rate region (blue area), Nash Bargaining Solution (the area
     of the white rectangular)
    (b) Per user price of anarchy for frequency selective Rayleigh fading channel.
  SNR=30 dB.}
  \label{Nash_graphs}
\end{figure}

 We can also give a geometrical interpretation to the solution. In Figure \ref{FDM_rate_region}
 we draw the feasible total rate that player $1$ can obtain as a function of the
total rate of player $2$. The enclosed area in blue, is the
achievable utilities set. Since, the frequency bins are sorted according to $L_k$ the
set is convex. The point $R_C=(R_{2C},R_{1C})=(10,15)$ is the point
of disagreement. If the point $R_C$ is inside the achievable
utility set there is a solution. The slope of the boundaries of the
achievable utilities set with respect to the $-x$ axis is $L_k$. The
vector $R_C-B$ connects the point $R_C$ and the point B is with the
same slope with respect the $x$ axis, this is the geometrical interpretation of (\ref{rule1}). The area of the white
rectangular is the value of Nash's product function.

The results can be generalized in several directions:

First, if the values $\left\{L(k): k=1,...,K\right\}$ are not all
distinct then if there is a solution one can always find allocation
such that at most a single frequency has to be shared.

Second, in the general case of $N$ players the optimization problem
has similar KKT conditions and can be solved using convex
optimization algorithm. Moreover, the optimal solution has at most
${K} \choose {2}$ frequencies that are shared between different
players. This suggests, that the optimal FDM NBS is very close to
the joint FDM/TDM solution. It is obtained by allocating the common
frequencies to one of the users.
Third, while the method described above fits well to stationary channels, the method is also useful
when only fading statistics is known. In this case the coding
strategy will change, and the achievable rate in the competitive
case and the cooperative case are given by \beq \bea{l} {\tilde
\mR_{nC}}\left(\vp_i\right)=
 \sum_{k=1}^{K}E\left[ \log_2\left(1+\frac{|h_{nn}(k)|^2p_n(k)}{\sum_{m \neq n}
 |h_{nm}(k)|^2 p_m(k)+\gs_n^2(k)}\right)\right]\\
{\tilde R_n(\vga_n)}=\sum_{k=1}^K \ga_{n}(k)E\left[\log_2
\left(1+\frac{|h_{nn}(k)|^2 p_n(k)}{\gs_n^2(k)}\right)\right], \ena
\eeq respectively. All the rest of the discussion is unchanged,
replacing $R_{nC}$ and $R_n(\vga_n)$ by ${\tilde \mR_{nC}},{\tilde
R_i(\vga_n)}$ respectively. This is particularly attractive, when
the computations are done in distributed way. {\em In this case only
channel state distributions are sent between the units}. Hence the time
scale for this data exchange are much longer. This implies that
method can be used without a central control, by exchange of
parameters between the units at a very low rate.

Fourth, computing the NBS under total power constraint is more difficult to
solve. Several ad-hoc techniques have been proposed in the literature.
Recently, it was shown that for this case there exist an algorithm
which can find the optimal solution \cite{zehavi2009}.

\section{Signal processing and computational issues}
\label{signal_processing}
The basic requirement behind any adaptive approach for spectrum management and co-existence of communication networks, is the capability to
adaptively shape the spectrum. While ten years ago this was beyond the capabilities of commercial communication systems, this is no longer
the situation. The signal processing hardware available at most of the modern chips is sufficient for this purpose, and the marginal cost of this
spectrum sensing and shaping capability is rapidly diminishing.
The main ingredients required to implement game theoretic approaches, are varying. Typically we distinguish three levels of management:
Autonomous, where each unit operated with no exchange of information with other units or management centers, distributed, where a management
center provides some parameters to the units, which then design the spectrum and transmit by themselves, centralized where all the spectrum design
is performed by a spectrum management center that collects information from units, and performs centralized processing. The competitive solutions
correspond to the autonomous operation. For these, each communication unit (player) is required to have a spectral analysis module, capable of estimating
channel transfer functions and noise spectrum. The regulated versions of the competitive games such as ASB, IPP and FM-IWF
belong to this second family of distributed solutions where the spectrum management center provides some a-priori network information or pricing functions
as well as a list of the desired rates.
For the distributed solutions, a low rate communication with a center or with adjacent units is
required like in the Nash bargaining solution of the previous section where only channel distributions are exchanged between users.

Finally, an important system and signal processing issue is synchronization of the various players. This is especially important for multi-carrier
systems, where inter channel interference can be severe when the various units are not locked to a common time and frequency reference.
\section{Applications}
\subsection{Weak interference: The DSL case}
\begin{figure}
    \begin{center}
    \subfigure[Simulation setup]
    {
    \label{IPP_setup}
    \includegraphics[width=0.35\textwidth]{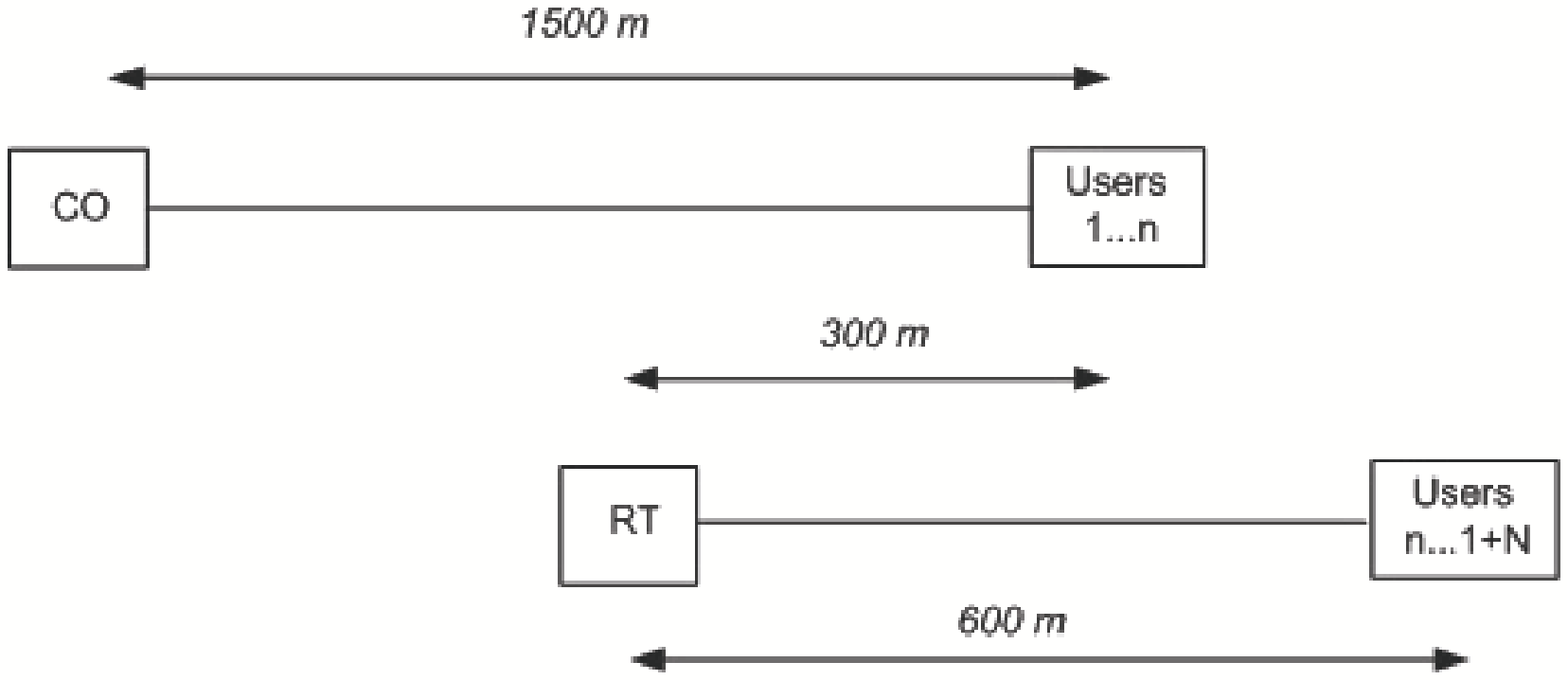}}
    \subfigure[Rate region]
    {
    \label{near_far_RR}
    \includegraphics[width=0.35\textwidth]{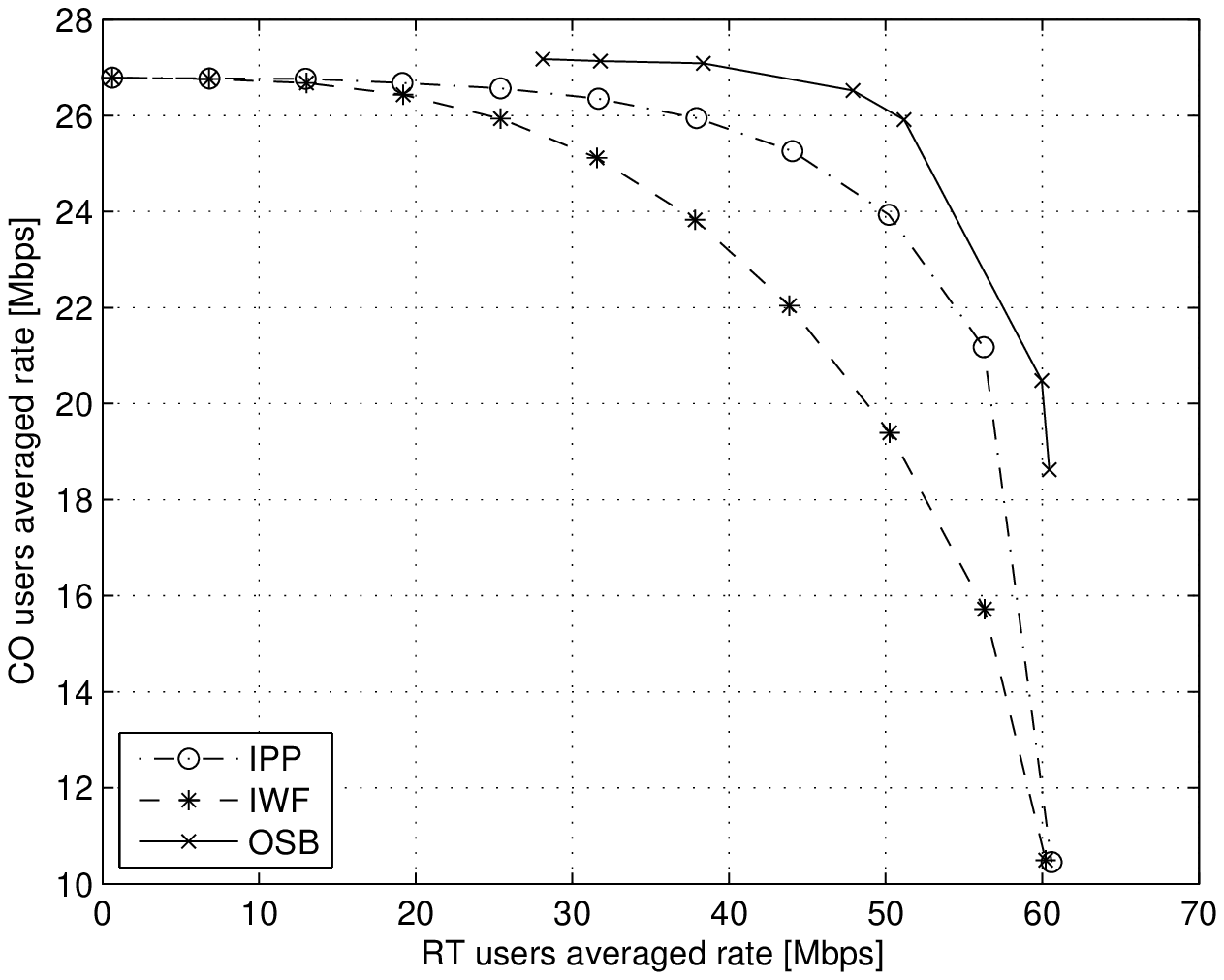}}
    \end{center}
    \caption{(a) Simulation setup. 3 users at each location. (b) Rate regions of FM
    IWF, IPP and OSB. The optimal OSB is centralized and computationally very expensive.
    It is given as a reference to the optimal performance.}
  \label{IPP}
\end{figure}

The DSL channel is an interesting example for testing algorithms emerging from game theoretic considerations. The iterative waterfilling
algorithm \cite{yu2002} has been successfully implemented for distributed spectrum coordination of DSL lines. However the drawbacks
caused by the prisoner's dilemma suggest that the strictly competitive approach (RA-IWF) is inappropriate for real life applications.
Several amendments have been proposed. The first is the fixed margin iterative waterfilling \cite{yu2002}. In this algorithm the players are provided
with a fixed target rate and each user, independently minimizes its total transmit power. As shown by Pang et al. \cite{pang_information_theory} this is a generalized Nash
game that converges if the interference is sufficiently weak. In \cite{noam2009} a generalization of the FM-IWF is proposed, that favors weak users
using a pricing mechanism termed iterative power pricing. This pricing mechanism improves the performance of the FM-IWF.
The game theoretic approaches have very good performance when compared to optimal spectrum management techniques, as shown in figure \ref{IPP}.
\subsection{Medium and strong interference - Wireless technologies}
The rapid adoption of wireless services by the public, have caused a
remarkable increase in demand for reliable high data rate Internet
access. This process motivated the development of new technologies.
New generation of cellular systems like LTE and WiMax operating in
the licensed band will be lunched in the near future. In the
unlicensed band, 802.11N with MIMO technology are going to become
part of our daily life.  The capacity of future wireless data
networks will inevitably be interference limited due, the the
limited radio spectrum. It is clear that any cooperation between the
different networks or base stations sharing the  same spectral
resource can offer significant improvement in the utilization of the
radio resources. Even in the same cell, cooperation between sectors
can improve the over all spectral efficiency (bit/Hz/sec./sector).
OFDMA technology is capable to allocate efficiently frequency bins
based on the channel response of the user.  In \cite{ZhuHJ}, a
noncooperative game approach was employed for distributed
sub-channel assignment, adaptive modulation, and power control for
multi-cell OFDM networks. The goal was to minimize the overall
transmitted power under maximal power and per user minimal rate
constraints. Based on simulation results, the proposed distributed
algorithm reduces the overall transmitted power in comparison with
pure water-filling scheme for a seven-cell case.  Kwon and Lee
\cite{Kwon} presented a distributed resource allocation algorithm
for multi-cell OFDMA systems relying on a noncooperative game in
which each base station tries to maximize the system performance
while minimizing the cochannel interference. They proved that there
exists a Nash equilibrium point for the noncooperative game and the
equilibrium is unique in some constrained environment. However, Nash
equilibrium achieved by the distributed algorithm may not be as
efficient as the resource allocation obtained through centralized
optimization. To demonstrate the
advantage of the Nash bargaining solution over competitive
approaches for a frequency selective interference channel we assume
that two users are sharing a frequency selective Rayleigh
fading channel. The direct channels have unit fading variance and SNR of 30 dB. The users suffer from
cross interference. The cross channels fading variance was varied from -10 dB to 0 dB ($\gs^2_{h_{ij}}=0.1,...1$).
The spectrum consisted of 32 parallel frequency bins with independent fading matrices. At each interference level
of interference $\gs_1^2=\gs_{h_{21}}^2,\gs_2^2=\gs_{h_{12}}^2$ we
randomly picked 25 channels (each comprising of 32 2x2 random matrices).
The results of the minimal relative improvement (\ref{min_delta})
are depicted in figure \ref{NBS}. \beq
 \label{min_delta}
 \gD_{\min}=\min\left\{ R^{NBS}_1 / R^C_1, R^{NBS}_2/R^C_2 \right\}.
 \eeq
The relative gain of the Nash bargaining solution over the
competitive solution is 1.5 to 3.5 times, which clearly demonstrates
the merits of the method.
%
%

\bibliographystyle{ieeetr}

\end{document}